\documentclass[a4paper,11pt]{article}
\usepackage{pos}
\usepackage{slashed,bbm,epsfig,color}

\def\mathswitch#1{\relax\ifmmode#1\else$#1$\fi}
\def\mathswitchr#1{\relax\ifmmode{\mathrm{#1}}\else$\mathrm{#1}$\fi}
\newcommand{\scrs}{\scriptscriptstyle}
\newcommand{\sw}{\mathswitch {s_{\scrs\mathswitchr W}}}
\newcommand{\cw}{\mathswitch {c_{\scrs\mathswitchr W}}}
\newcommand{\msbar}{\mathswitch{\overline{\mathswitchr{MS}}}}
\newcommand{\seff}[1]{\sin^2\theta_{\rm eff}^{#1}}

\title{%TASI 2020 Lectures on 
Precision Tests of the Standard Model}
\author{Ayres Freitas}

\affiliation{PITT PACC, Department of Physics \& Astronomy, University of Pittburgh\\
Pittsburgh, PA 15260, USA}

\emailAdd{afreitas@pitt.edu}

\abstract{%
This write-up of lectures given at TASI 2020 provides an introduction into
precision tests of the electroweak Standard Model. The lecture notes begin with
a hands-on review of the (on-shell) renormalization procedure, and subsequently
highlight a few subtleties that occur in the renormalization of a theory with
electroweak symmetry breaking and massive gauge bosons. After that a set of
typical electroweak precision observables is introduced, as well as a range of
input parameter measurements that are needed for making predictions within the
Standard Model. Finally, it is discussed how comparisons of the electroweak
precision observables between experiment and theory can be used to stress-test
the Standard Model and probe new physics.
}

\FullConference{%
  Theoretical Advanced Study Institute: The Obscure Universe: Neutrinos and Other Dark Matters - TASI2020\\
  1-26 June, 2020\\
  Boulder, Colorado, USA
}

\begin{document}
\maketitle

\section{Introduction}

The Standard Model of electroweak interactions
\cite{Glashow:1961tr,Weinberg:1967tq,Salam:1968rm} is at the core of today's
understanding of fundamental physics. The breaking of the electroweak symmetry
through the Higgs mechanism is the origin of the masses of all other elementary
particles in the Standard Model (SM), and it explains the apparent ``weakness''
of the weak interactions in low-energy physics. In contrast to the strong
interactions, one can make reliable high-precision predictions using
perturbation theory for electroweak observables. The realization that the
electroweak theory is perturbatively calculable \cite{tHooft:1972tcz} has tremendously
advanced the understanding of its theoretical structure and provided the
opportunity for precise experimental tests of all its aspects.

Through comparisons of precision measurements of properties of the electroweak
gauge bosons with theoretical predictions within the SM in the 1990s and 2000s, 
it was possible to put constraints on the some of the last undiscovered
components of the SM: the top quark and the Higgs boson (see Figs.~1.16, 8.3,
8.11, 8.13 in Ref.~\cite{ALEPH:2005ab}). At the same
time, electroweak precision tests put important constraints on physics beyond
the SM and have conclusively ruled out some models. These lectures provide an
introduction into the most common electroweak precision observables, their
theoretical underpinnings, and how they can be used to test the SM and physics
beyond the SM.

It is assumed that the reader is familiar with the general structure of the
Standard Model and general aspects of quantum field theory, such as Lagrangians,
Feynman rules, perturbation theory, gauge symmetries and Ward identities, and
electroweak symmetry breaking through the Higgs mechanism. Good examples for
pedagogical reviews of the foundations of the Standard Model can be found in
Refs.~\cite{Novaes:1999yn,Peskin:2017emn,Arbuzov:2018fza}.

Since the topic of these lectures requires a solid understanding of foundational
aspects of higher-order corrections and renormalization, they begin with a
review of renormalization in QED and in the Standard Model in
section~\ref{renorm}. Section~\ref{ewpos} discusses a range of quantities known
as electroweak precision observables, which play an important role in detailed
tests of the Standard Model, in particular its electroweak symmetry breaking
sector. Finally, in section~\ref{bsm}, it is shown how electroweak precision
observables can be used to probe and constrain physics beyond the Standard
Model, with an emphasis on models of neutrino physics and dark matter, owing to
the themse of the TASI 2020 school.

Throughout this document, the following conventions for the metric tensor and
Dirac algebra are being used:
\begin{align}
(g_{\mu\nu}) &= \text{diag}(+1,-1,-1,-1), &
\{\gamma_\mu,\gamma_\nu\} &= 2g_{\mu\nu}\,\mathbbm{1}_{4{\times}4}, &
\{\gamma_\mu,\gamma_5\} &= 0. \label{conv}
\end{align}
The document also contains a handful of exercise problems that the reader is
encouraged to try to solve. Answers to the problems are given at the very end of
the document.

%%%%%%%%%%%%%%%%%%%%%%%%%%%%%%%%%%%%%%%%%%%%%%%%%%%%%%%%%%%%%%%%%%%%%%%%%%%%%%%

\section{Renormalization}
\label{renorm}

%%%%%%%%%%%%%%%%%%%%%%%%%%%%%%%%%%%%%%%%%%%%%%%%%%%%%%%%%%%%%%%%%%%%%%%%%%%%%%%

\subsection{Renormalization in QED}

Before discussing renormalization in the Standard Model (SM), let us first
illustrate the main concepts for a simplet theory: Quantum Electrodynamics
(QED), which describes a charged Dirac fermion\footnote{The extension to several
fermions with different charges and masses is straightforward.} 
$\psi$ that interacts
with the photon field $A_\mu$. Its Lagrangian is given by
\begin{align}
{\cal L} &= -\tfrac{1}{4} F_{0,\mu\nu} F_0^{\mu\nu} + \overline{\psi}_0
\bigl(i\slashed{\partial} + e_0 \slashed{A}_0 - m_0 \bigr) \psi_0,
&
F_{0,\mu\nu} &= \partial_\mu A_{0,\nu} - \partial_\nu A_{0,\mu}.
\end{align}
This expression contains two free parameters: $e_0$ and $m_0$, the charge and
mass of the fermion $\psi_0$, respectively.

When including radiative corrections, these parameters will in general differ
from the observable charge and mass of the fermion. Denoting the latter as $e$
and $m$, the relation can be written as
\begin{align}
e_0 &= Z_e\,e = (1+\delta Z_e)e,
&
m_0 &= m + \delta m
\end{align}
The quantities $\delta X$ are called \emph{counterterms}. Here and in the
following, the index ``0'' is used for Lagrangian (``bare'') quantities, whereas
the corresponding symbols without subscript denote physical (renormalizated)
quantities.

To determine the counterterms, one needs to specify a set of
\emph{renormalization conditions} that define what we mean by ``physical
quantities.'' For the charge and mass, we can find a set of conditions that
formally reflect how these quantities are typically measured in an experiment:

\paragraph{Mass \boldmath $m$:} The physical mass is defined as the pole in the fermion
propagator
\begin{align}
D(p) \equiv \frac{i}{\slashed{p}-m} = \frac{i(\slashed{p}+m)}{p^2-m^2},
\label{propf}
\end{align}
since the peak in the propagation probability $|D(p)|^2$ for $p^2=m^2$
corresponds to long-distance propagation ($i.\,e.$ an actual observable
particle).

When computing the propagator from the Lagrangian, one must
include radiative corrections, leading to
\begin{align}
&\raisebox{-1em}{\epsfig{figure=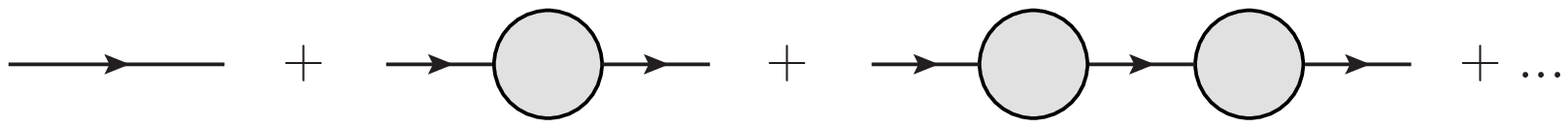, height=2em, bb=74 682 540 716,
 clip=true}} \label{dysonf1} \\
= &\,\frac{i}{\slashed{p}-m_0}
+ \frac{i}{\slashed{p}-m_0} i\Sigma(p) \frac{i}{\slashed{p}-m_0}
+ \frac{i}{\slashed{p}-m_0} i\Sigma(p) \frac{i}{\slashed{p}-m_0}
 i\Sigma(p) \frac{i}{\slashed{p}-m_0}
+ ... \label{dysonf2} \\
= &\,\frac{i}{\slashed{p}-m_0+\Sigma(p)}\label{dysonf3} 
\end{align}
Here $\Sigma(p)$ is the \emph{self-energy} of the fermion, which represents all
one-particle irreducible loop diagrams contributing to the fermion two-point
functions [depicted symbolically by the blob in \eqref{dysonf1}].
Eq.~\eqref{dysonf2} is called a \emph{Dyson series}, which can be resummed as a
geometric series, leading to eq.~\eqref{dysonf3}.

$\Sigma(p)$ can contain $\gamma$ matrices and thus can be expanded as a sum of
the following terms:
\begin{align}
\Sigma(p) &= \Sigma_S(p^2) + \gamma_\mu p^\mu \, \Sigma_V(p^2)
 + \underbrace{\overbrace{\gamma_\mu p^\mu \gamma_\nu p^\nu}^{=p^2} \,
 \Sigma_T(p^2)}_{\to \text{ absorb in }\Sigma_S} + ...
\end{align}
Owing to Lorentz invariance, the coefficients $\Sigma_X$ can only depend on
$p^2$. The term linear in $\gamma_\mu$ (called the ``vector'' part of the
self-energy) must be proportional to $p^\mu$ since this is the only other
4-vector that can be contracted with $\gamma_\mu$. The term with two gamma
matrices (the ``tensor'' part) can be rewritten, using \eqref{conv}, as being
proportional to $p^2$ and thus it is already captured by the ``scalar'' part
$\Sigma_S$. In the same way, all terms with three or more gamma matrices can be
absorbed into $\Sigma_S$ and $\Sigma_V$.

Demanding that the propagator has a pole for $p^2=m^2$, or equivalently
$\slashed{p}=m$ (see eq.~\eqref{propf}) leads to the condition
\begin{align}
&0 = \underbrace{\slashed{p} - (m}_0 + \delta m) + \bigl[\slashed{p}
 \Sigma_V(p^2)+ \Sigma_S(p^2)\bigr]_{p^2=m^2,\slashed{p}=m} \\
\Rightarrow\quad &\delta m = m\,\Sigma_V(m^2) + \Sigma_S(m^2)
\end{align}

\medskip
\noindent
\begin{minipage}[b]{.75\textwidth}
\paragraph{Charge \boldmath $e$:} The physical charge is defined as the strength
of the electromagnetic coupling in the Thomson limit: an on-shell fermion
($p_1^2=p_2^2=m^2$) interacts with a static electric field ($i.\,e.$ a photon
with zero momentum, $k = p_2-p_1 \to 0$). Denoting the sum of all vertex
diagrams by $\Gamma_\mu(p_1,p_2)$, this means that
\begin{align}
\bar{u}(p_1)\, i\Gamma_\mu(p_1,p_1)\, u(p_1) = \bar{u}(p_1)\,
 ie\gamma_\mu\, u(p_1) \quad \text{for } p_1^2=m^2 \label{chcond}
\end{align}
\end{minipage}
\hfill\raisebox{2em}{%
\epsfig{figure=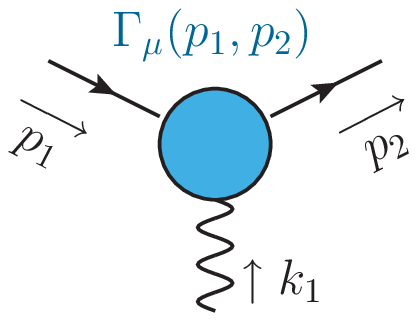, width=.2\textwidth, bb=235 620 355 710, clip=}}

\vspace{1ex}\noindent
The vertex factor can be written as a tree-level piece and a term
$\delta\Gamma_\mu$ that subsumes all loop contributions. The latter can be
related to the fermion self-energies using the QED Ward identity, which implies
that
\begin{align}
k^\mu\,\delta\Gamma(p,p+k) &= e\bigl[\Sigma(p+k)-\Sigma(p)\bigr] \\
\Rightarrow\quad \delta\Gamma(p,p) &= e\, \lim_{k\to 0} \frac{1}{k^\mu}
 \bigl[\Sigma(p+k)-\Sigma(p)\bigr] = e\, \frac{\partial}{\partial p^\mu}
 \Sigma(p) \label{verward}
\end{align}

\medskip
\noindent
\begin{minipage}[b]{.75\textwidth}
\paragraph{Field renormalization:} Until now, we only considered the
renormalization of the parameters in the Lagrangian. However, the fields
themselves receive quantum corrections, due to self-energy contributions in the
external legs of any physics process (also called ``wave function''
renormalization). These corrections can be absorbed by redefining the fields:
\begin{align}
A_0^\mu &= \sqrt{Z_A} \, A^\mu, &
\psi_0 &= \sqrt{Z_\psi} \, \psi,
\end{align}
\end{minipage}
\hfill\raisebox{2em}{%
\epsfig{figure=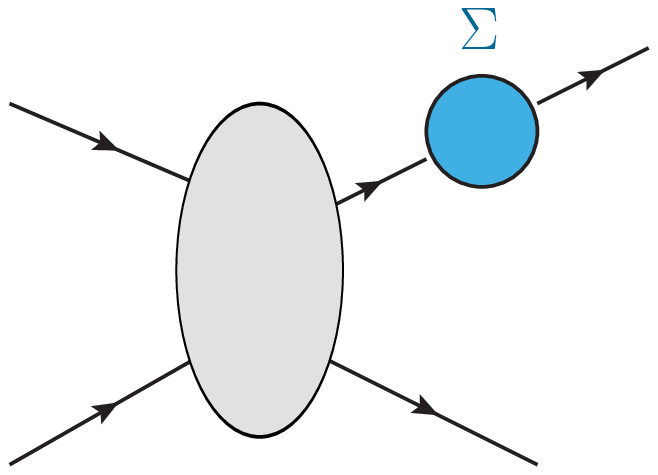, width=.2\textwidth, bb=205 574 390 710, clip=}}

\smallskip\noindent
where, as before, we can write $Z_X = 1 + \delta Z_X$, where $\delta Z_X$ is the
counterterm due to loop corrections. These counterterms should cancel the
self-energy corrections on the external legs. For an external fermion, one
therefore should demand
\begin{align}
&\bigl[Z_\psi(\slashed{p}+m_0) + \Sigma(p)\bigr]_{\slashed{p}\to m} =
 (\slashed{p}+m)_{\slashed{p}\to m}
\intertext{Taking the derivative $p^\mu \frac{\partial}{\partial p^\mu}$ on both
sides yields}
&\Bigl[p^\mu (1+\delta Z_\psi)\gamma_\mu + p^\mu \frac{\partial}{\partial p^\mu}
\Sigma(p)\Bigr]_{\slashed{p}\to m} = p^\mu\gamma_\mu\big|_{\slashed{p}\to m} 
 \label{psi1} \\
\Rightarrow\quad &\delta Z_\psi = -\frac{p^\mu}{m}\; 
 \frac{\partial}{\partial p^\mu}\Sigma(p)\Bigr|_{\slashed{p}\to m}
= -\Sigma_V(m^2) -2m \Sigma'_V(m^2) - 2 \Sigma'_S(m^2)
\end{align}
For external photons, we need to use the photon self-energy,
$\Sigma_{\mu\nu}(k)$, which can be decomposed into a transverse and a
longitudinal part:
\begin{align}
\Sigma_{\mu\nu}(k) &= \Bigl(g_{\mu\nu} - \frac{k_\mu k_\nu}{k^2}\Bigr )
\Sigma_T(k^2) + \frac{k_\mu k_\nu}{k^2}\Sigma_L(k^2)
\end{align}
Invoking the QED Ward identity, $k^\mu \Sigma_{\mu\nu} =0$ immediately tells us
that $\Sigma_L = 0$.

Applying the Dyson summation to the remaining transverse part of the self-energy
yields the following result for the photon propagator (in Feynman gauge):
\begin{align}
\Bigl(g_{\mu\nu} - \frac{k_\mu k_\nu}{k^2}\Bigr )\frac{-i}{k^2+\Sigma_T(k^2)}
 + \frac{k_\mu k_\nu}{k^2}\,\frac{-i}{k^2} \label{propa}
\end{align}
Note that the last term in this equation changes if a different gauge than
Feynman gauge is adopted. Including the field renormalization counterterm
requires to modify \eqref{propa} according to $k^2+\Sigma_T(k^2) \to Z_A
k^2+\Sigma_T(k^2)$. Demanding that $Z_A$ should compensate the self-energy
contribution for an on-shell photon leads to
\begin{align}
&\bigl[ Z_A k^2 + \Sigma_T(k^2) \Bigr]_{k^2 \to 0} = k^2\big|_{k^2\to 0} \\
\Rightarrow\quad &\delta Z_A = -\Sigma'_T(0)
\end{align}

\paragraph{Field renormalization effects in charge renormalization:}
For the proper evaluation of the charge renormalization condition
\eqref{chcond}, we must include the field renormalization factors, yielding
\begin{align}
\sqrt{Z_A} \,Z_\psi\, e_0\, \gamma_\mu + \delta \Gamma_\mu = e\gamma_\mu
\end{align}
Expanding the left-hand term to leading order in perturbation theory yields
$\sqrt{Z_A} \,Z_\psi\, e_0 = e(1+\delta Z_e + \frac{1}{2}\delta Z_A + \delta
Z_\psi + ...)$. Furthermore we can use that $\delta\Gamma = e\frac{\partial}{\partial
p^\mu}\Sigma$ according to \eqref{verward} and $\frac{\partial}{\partial
p^\mu}\Sigma = \delta Z_\psi \gamma_\mu$ according to \eqref{psi1}. Thus one
obtains a rather simple result for the charge counterterm:
\begin{align}
&\delta Z_e = -\tfrac{1}{2}\delta Z_A && \text{[at 1-loop order]} \label{dze1}
\end{align}
An explicit one-loop calcution of the fermion loop diagram below yields
\begin{align}
&\raisebox{-1.2em}{\epsfig{figure=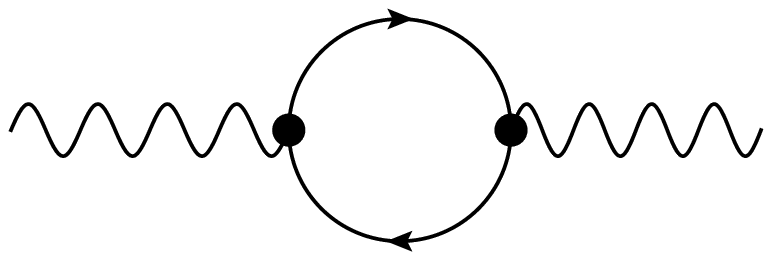, height=2.4em}}
&\Sigma'_T(0) &= \frac{\alpha}{3\pi}\biggl(\frac{2}{4-d}-\gamma_{\rm E} -
 \ln \frac{m^2}{4\pi\mu^2}\biggr) \label{se1}
\end{align}
where $d$ and $\mu$ are the number of space-time dimensions and the
regularization scale in dimensional regularization, respectively. Furthermore,
$\gamma_{\rm E} \approx 0.577216$ is Euler's constant.

\medskip\noindent
Extending the QED theory to include all SM fermions, this becomes
\begin{align}
\Sigma'_T(0) &= \sum_f N_{\rm c}^f Q_f^2\, \frac{\alpha}{3\pi}\biggl(\frac{2}{4-d}-\gamma_{\rm E} -
 \ln \frac{m_f^2}{4\pi\mu^2}\biggr) \label{sigmap}
\end{align}
where $N_{\rm c}^f=1\,(3)$ for leptons (quarks) and $Q_f$ is the electric charge
of the fermion species $f$ in units of the positron charge $e$. A problem with
\eqref{sigmap} is the fact that light quark masses ($m_u,\,m_d,\,m_s$) are
ill-defined, since QCD at the scale $m_{u,d,s}$ is inherently non-perturbative,
and thus a perturbative calculation as in eq.~\eqref{sigmap} is not adequate.
\label{deltaalpha}

This problem can be circumvented by using a \emph{dispersion relation} that
establishes a relationship between $\Sigma'_T(0)$ and the process $e^+e^- \to
\text{hadrons}$, which can be obtained from data. In order to so, as a first
step we will rewrite $\Sigma'_T(0)$ as follows:
\begin{align}
\Sigma'_T(0) &= \Pi(0) = \underbrace{\Pi(0) -
\text{Re}\,\Pi(M_Z^2)}_{\equiv\Delta\alpha} + 
 \text{Re}\,\Pi(M_Z^2), \qquad \Pi(Q^2) \equiv \frac{\Sigma_T(Q^2)}{Q^2}
\label{delal1}
\end{align}
Here the term $\Delta\alpha$ is UV finite, while $\Pi(M_Z^2)$ depends on the
light quark masses only through powers of $m_q^2/M_Z^2 \approx 0$ and thus can
be computed perturbatively to very good accuracy. The choice of $M_Z$ for the
separation scale in \eqref{delal1} is somewhat arbitrary; the only requirement
is that this scale should be much larger than $\Lambda_{\rm QCD}$. However,
$M_Z$ has become the conventional choice in the literature.

Furthermore, $\Delta\alpha$ can be divided into a leptonic and a hadronic part,
$\Delta\alpha = \Delta\alpha_{\rm lept} + \Delta\alpha_{\rm had}$, where
$\Delta\alpha_{\rm lept}$ can also be reliably calculated using perturbation
theory \cite{Steinhauser:1998rq,Sturm:2013uka}.
On the other hand, $\Delta\alpha_{\rm had}$ can be related to the process $e^+e^- \to
\text{hadrons}$ using a dispersion integral (see below for the derivation):
\begin{align}
\Delta\alpha_{\rm had} &= -\frac{\alpha}{3\pi}\int_0^\infty ds' \;
\frac{R(s')}{s'(s'-M_Z^2-i\epsilon}, &
R(s) &= \frac{\sigma[e^+e^- \to \text{hadrons}]}{\sigma[e^+e^- \to \mu^+\mu^-]}
\label{delal2}
\end{align}
For $s \lesssim 2$~GeV, $R(s)$ is typically extracted from data collected at 
several $e^+e^-$ colliders, while QCD perturbation theory can be used for $s
\gtrsim 2$~GeV. In many analyses, data is also used near the $c\bar{c}$ and
$b\bar{b}$ thresholds, although it has been argued that perturbation theory can
also be used in these regions \cite{Erler:1998sy,Erler:2017knj}. For recent
evaluations of $\Delta\alpha_{\rm had}$ 
from $R(s)$, see Refs.~\cite{Blondel:2019vdq,Davier:2019can,Keshavarzi:2019abf}.

Efforts are also underway to compute $\Delta\alpha_{\rm had}(s) \equiv \Pi_{\rm had}(0) -
\text{Re}\,\Pi_{\rm had}(s)$ using lattice QCD \cite{Burger:2015lqa,Ce:2019imp}, but more
work and a more detailed evaluation of systematic errors will be needed before
they can be applied in phenomenological applications. Finally, it is possible to
extract $\Delta\alpha(s)$ directly from measurements of Bhabha scattering
\cite{Abbiendi:2005rx,Achard:2005it,KLOE-2:2016mgi}, but
the currently achievable precision is not competitive with the dispersion relation
method.

%------------------------------------------------------------------------------
\begin{figure}
\centering
\epsfig{figure=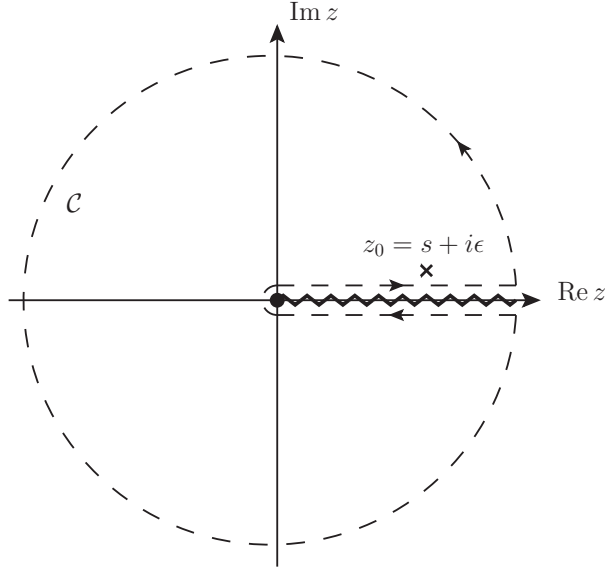, width=8cm, bb=130 400 456 706}
\caption{Integration contour for using Cauchy's integral theorem for a function
that has a branch cut along the positive real axis (indicted by a zigzag line).
The circle section is understood to have a radius $R \to \infty$.}
\label{disp1}
\end{figure}
%------------------------------------------------------------------------------

\begin{quote}
\textbf{Derivation of eq.~\eqref{delal2}:} Suppose a function $f(z),\,z\in
\mathbbm{C}$ has a branch cut along the positive real axis, but is analytical
elsewhere. One can then apply Cauchy's integral theorem for a contour $\cal C$
that excludes the branch cut, see Fig.~\ref{disp1}
\begin{align}
f(z_0) = \frac{1}{2\pi i} \oint_{\cal C} dz' \; \frac{f(z')}{z'-z_0}
\end{align}
If $f(z)$ vanishes sufficiently fast for $|z| \to\infty$, only the parts of
$\cal C$ along the real axis need to be considered. For $z_0=s+i\epsilon$ one
then obtains
\begin{align}
f(s+i\epsilon) = \frac{1}{2\pi i} \int_0^\infty ds' \; \frac{f(s'+i\delta) -
f(s'-i\delta)}{s'-s-i\epsilon}
\end{align}
where $\delta < \epsilon$ are both infinitesimally small. Applying this to
$f(z)=\Pi(z)$ and noting that $\Pi(s'-i\delta) = \Pi^*(s'+i\delta)$,
\begin{align}
\text{Re}\,\Pi(s) = \frac{1}{\pi} \int_0^\infty ds' \; \frac{\text{Im}\,
 \Pi(s'+i\delta)}{s'-s-i\epsilon} \label{disp2}
\end{align}
[The $i\epsilon$ on the l.h.s.\ can be dropped if we only consider the real part
of $\Pi$.]

Now we can relative the photon vacuum polarization $\Pi(s)$ to the matrix
element for $e^+e^- \to e^+e^-$ with a photon self-energy in the s-channel:
\begin{align}
\text{Im}\, \Pi(s') &= \frac{1}{e^2} \;\; \text{Im}\;{\cal M}\biggl\{
 \raisebox{-.7em}{\epsfig{figure=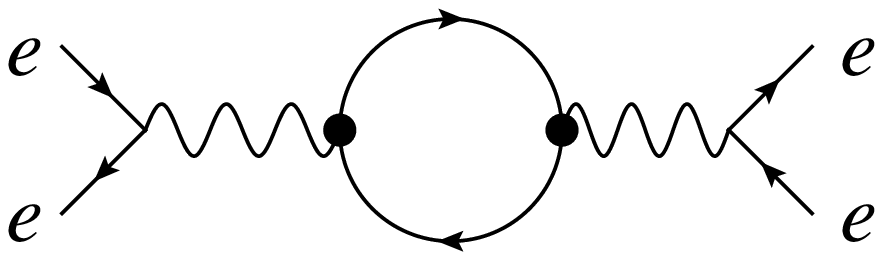, height=2em}}\biggr\}_{\theta=0} \\
&= \frac{s'}{e^2} \sum_f \sigma[e^+e^- \to f\bar{f}] && \text{[optical theorem]}
 \label{opt1}\\
&= \frac{s'}{e^2} \,R(s')\,\underbrace{\sigma[e^+e^- \to
\mu^+\mu^-]}_{4\pi\alpha^2/(3s')} \label{opt2}
\end{align}
where ``$\theta=0$'' indicates that we are restricting ourselves to forward
scattering, $i.\,e.$ the kinematics of the final-state $e^+e^-$ are the same as
in the initial state. Then we can apply the optical theorem in \eqref{opt1}.
Inserting \eqref{opt2} into \eqref{disp2}, one arrives at
\begin{align}
\text{Re}\;\bigl[\Pi(s)-\Pi(0)\bigr] =
\frac{\alpha}{3\pi} \int_0^\infty ds' \; R(s') \biggl[ \frac{1}{s'-s-i\epsilon}
- \frac{1}{s'}\biggr]
\end{align}
which immediately leads to the formula for $\Delta\alpha$ in \eqref{delal2}.
\rule{1ex}{1ex}

\end{quote}

\paragraph{Exercise:} Compute the result in \eqref{se1}. {\sl Hint:} 
$\Sigma_T(k^2)$ can be computed with Feynman rules and standard techniques, with
the result $\frac{\alpha}{3\pi}\Bigl[ 
 \frac{3(d/2-1)k^2+6m^2}{d-1}B_0(k^2,m^2,m^2) - \frac{4(d-2)}{d-1} A_0(m^2)
 \Bigr]$. To compute the derivative of $B_0(k^2,m^2,m^2)$,
show that $\frac{\partial^2}{\partial k_\mu \partial k^\mu} f(k^2)
 = 4k^2\,f''(k^2) + 2d\,f'(k^2)$. Then apply $\frac{\partial^2}{\partial k_\mu
 \partial k^\mu}$ \emph{inside} the integral and use this to compute
 $\frac{\partial}{\partial(k^2)}B_0(k^2,m^2,m^2)\big|_{k^2=0}$. Finally, express the
 result in terms of $A_0(m^2)$ and derivatives thereof and use $A_0(m^2) =
 m^2\Bigl[\frac{2}{4-d} - \gamma_E - \ln\frac{m^2}{4\pi\mu^2} + 1\Bigr].$
\label{se1a}

%%%%%%%%%%%%%%%%%%%%%%%%%%%%%%%%%%%%%%%%%%%%%%%%%%%%%%%%%%%%%%%%%%%%%%%%%%%%%%%

\subsection{On-shell Renormalization in the Standard Model}

In this subsection, the renormalization procedures from QED are extended to the
full Standard Model (SM). Some unique aspects related to electroweak symmetry
breaking and massive gauge bosons are reviewed in detail, whereas the remainig
aspects that are conceptually similar to QED are only summarized
briefly\footnote{A more detailed exposition of renormalization in the SM can be
found $e.\,g.$ in Ref.~\cite{Denner:1991kt}.}.

%------------------------------------------------------------------------------
\begin{table}
\centering
\renewcommand{\arraystretch}{1.4}
\begin{tabular}{|l|l|}
\hline
Gluon & $G_0^a = \sqrt{1+\delta Z_G}\,G^a \quad [a=1,...8]$ \\
Charged $W^\pm$ bosons & $W_0^\pm = \sqrt{1+\delta Z_W}\,W^\pm$ \\
%\hline
Photon, $Z$ boson & $\begin{pmatrix} Z_0 \\[-.5ex] A_0 \end{pmatrix}
 = \begin{pmatrix} \sqrt{1+\delta Z_{ZZ}} & \frac{1}{2}\delta Z_{AZ} \\[-.5ex]
   \frac{1}{2}\delta Z_{AZ} & \sqrt{1+\delta Z_{AA}} \end{pmatrix}
   \begin{pmatrix} Z \\[-.5ex] A \end{pmatrix}$ \\
\hline
Higgs boson & $H_0 = \sqrt{1+\delta Z_H}\,H$ \\
\hline
Fermions & $\begin{matrix} \psi^L_{f,0} = \sqrt{1+\delta Z^L_f}\,\psi^L_f
\\
\psi^R_{f,0} = \sqrt{1+\delta Z^R_f}\,\psi^R_f \end{matrix}$ \\
\hline
\end{tabular}
\caption{Field renormalization counterterms of the SM fields.}
\label{smfield}
\end{table}
%------------------------------------------------------------------------------

The field content of the SM and the associated field renormalization
counterterms are listed in Tab.~\ref{smfield}. The photon and $Z$ boson receive
a matrix-valued renormalization factor to account for mixing between these two
fields. Since the left- and right-handed fermions in the SM have different
interactions, the also receive independent renormalization factors.

In addition to the field renormalization terms, the SM also contains several
parameters that in general will be renormalized by higher-order effects:
\begin{itemize}
\item Gauge couplings $g,g',g_{\rm s}$ associated with the U(1), SU(2) and SU(3)
gauge interactions.
\item Yukawa couplings $y_f$. In general these are matrices in the space of the
three SM fermion generations, but for the purpose of these lecture we will
ignore CKM mixing\footnote{This approximation is justified by the fact that for
electroweak physics CKM mixing is most relevant in the third generation, due to
the enhancement from the large top-quark mass, but the CKM matrix is very nearly
unitary in the third row.}.
\item Higgs vacuum expectation value (vev) $v = \langle \phi_2 \rangle \approx
246$~GeV, where $\phi$ is the Higgs SU(2) doublet, and the Higgs self-coupling
$\lambda$.
\end{itemize}
For the renormalizion procedure, it is desirable to relate these parameters to
observables, such as
\begin{itemize}
\item the positron charge $e$ (in the Thomson limit);
\item the massive boson masses $M_W,\,M_Z,\,M_H$;
\item the fermion masses $m_f$ ($f=e,\mu,\tau,u,d,s,c,b,t$)\footnote{Neutrino
masses are exactly zero in the SM, in obvious conflict with observations.
However, the tiny neutrino masses are irrelevant for electroweak physics.}.
\end{itemize}
At tree-level, the relationship between parameters and these observables is
given by the following equations
\begin{flalign}
\hspace{3ex}
\begin{aligned}
&\bullet \; \cw \equiv \cos\theta_{\rm W} = \frac{M_W}{M_Z}, \quad \sw^2 = 1-\cw^2, \\
&\bullet \;g = \frac{e}{\sw}, \quad g'=\frac{e}{\cw}, \\
&\bullet \;v = 2M_W/g, \\
&\bullet \;\lambda = M_H^2/(2v^2). 
\end{aligned} && \label{osr}
\end{flalign}
Here the weak mixing angle $\theta_{\rm W}$ has been introduced for convenience.

The \emph{on-shell (OS) renormalization scheme} is defined by enforcing relations in
eq.~\eqref{osr} to all orders in perturbation theory.

Note the absence of $g_{\rm s}$ in this list. Since the strong coupling becomes
non-perturbative at low energies, there is no OS definition for $g_{\rm s}$.
Instead, the most common prescription for this coupling is the so-called \msbar\
scheme, where the counterterm is defined as a pure UV-divergent term in
dimensional regularization: \label{msbar1}
\begin{align}
\delta g_{\rm s} &= (4\pi e^{-\gamma_{\rm E}})^{L\varepsilon}
\Bigl(\frac{C_L}{\varepsilon^L} + \frac{C_{L-1}}{\varepsilon^{L-1}} +
... + \frac{C_1}{\varepsilon} \Bigr),
&& \varepsilon = \frac{2}{4-d}, \quad L = \text{loop order}
\end{align}
where the $C_i$ are chosen such that the sum of the $L$-loop corrections to the
$gq\bar{q}$ vertex and the vertex counterterm are UV-finite:
\begin{align}
\left[\raisebox{-2em}{\epsfig{figure=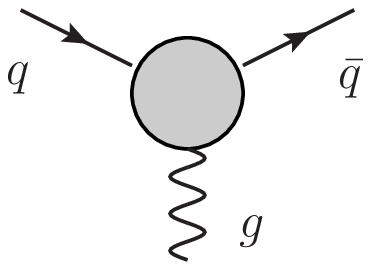, height=4.5em, 
 bb=240 640 350 720, clip=}}
\right]_{L\rm -loop}
+ \raisebox{-2em}{\epsfig{figure=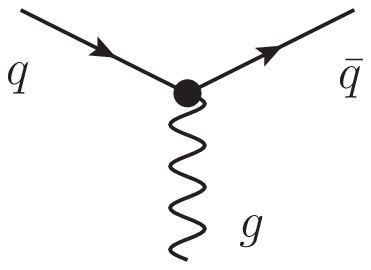, height=4.5em, 
 bb=240 640 350 720, clip=}} \times
\bigl[\sqrt{1+\delta Z_G}\,(1+\delta Z_q)(1+\delta g_{\rm s})\bigr]_{L\rm -loop}
= \text{finite}
\end{align}

\noindent
\paragraph{\boldmath $\gamma$--$Z$ mixing:}
The derivation of the OS counterterms proceeds in a similar fashion as for QED,
with a few modifications. For example, the expression for the charge counterterm
in \eqref{dze1} must be adjusted to account for photon--$Z$ mixing:
\begin{align}
\delta Z_{e(1)} &= -\tfrac{1}{2}\delta Z_{AA(1)} - \frac{\sw}{2\cw}
 \delta Z_{ZA(1)}
\end{align}
Here and the in the following the subcript $(n)$ indicates the loop order.

Additional renormalization conditions are needed to fix the mixing
counterterms $\delta Z_{ZA}$ and $\delta Z_{AZ}$. Within the OS scheme, this is
achieved by demanding that an on-shell photon does not mix with the $Z$ boson,
and conversly an on-shell $Z$ boson does not mix with the photon. At one-loop
order, we can write the photon--$Z$ two-point function as
\begin{align}
G_{\mu\nu(1)}^{AZ} &\equiv \;\raisebox{-1.3em}{\epsfig{figure=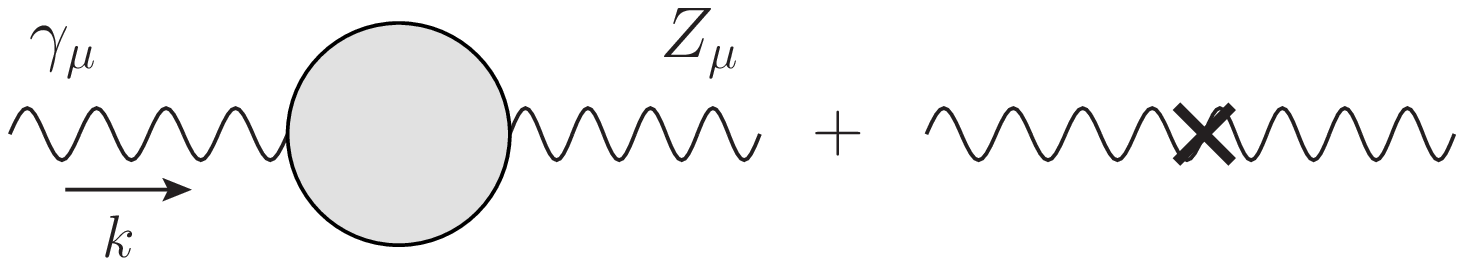, height=3em, bb=90 640 510
720, clip=}} \notag \\
& = -i\Bigl(g_{\mu\nu} - \frac{k_\mu k_\nu}{k^2} \Bigr)
\bigl[ \Sigma^{AZ}_{T(1)}(k^2) + k^2\tfrac{1}{2}\delta Z_{AZ(1)} +
(k^2-M_Z^2) \tfrac{1}{2} \delta Z_{ZA(1)} \bigr ] 
-i\frac{k_\mu k_\nu}{k^2} \bigl[ ... \bigr] \label{azmix} 
\end{align}
where the blob symbolizes all loop diagrams contributing at the given order,
whereas the cross symbolizes the counterterm contributions. The former is
formally described by the photon--$Z$ self-energy $\Sigma^{AZ}_{T(1)}$, whereas
the latter receives two contributions: the one with $\delta Z_{AZ(1)}$ stems
from the $A_0$ propagator, when the $A_0$ field then gets renormalized according to
Tab.~\ref{smfield}, while the one with $\delta Z_{ZA(1)}$ stems from the $Z_0$
propagator, when the $Z_0$ field gets renormalized. The
longitudinal part has not been spelled out in \eqref{azmix} because it does not contribute to
physical in- and out-states. Now, imposing the on-shell non-mixing conditions,
one obtains
\begin{align}
&G_{\mu\nu(1)}^{AZ} = 0 \quad \text{for } k^2 = 0
& \Rightarrow \quad \delta Z_{ZA(1)} &= 2\frac{\Sigma^{AZ}_{T(1)}(0)}{M_Z^2}
\\
&G_{\mu\nu(1)}^{AZ} = 0 \quad \text{for } k^2 = M_Z^2
& \Rightarrow \quad \delta Z_{AZ(1)} &=
-2\frac{\text{Re}\,\Sigma^{AZ}_{T(1)}(M_Z^2)}{M_Z^2}
\end{align}

\noindent
\paragraph{Unstable particles:} \label{unstable}
An additional complication arises for the renormalization of unstable particles,
since their self-energy has an imaginary part, $\text{Im}\,\Sigma(M^2)>0$, so
that the pole of the propagator becomes complex! In practice, in the SM, this is
relevant for the $W$ and $Z$ bosons and the top quark, since the width of all
other SM particles is negligibly small.

A detailed review of this issue can be found, $e.\,g.$, in
Ref.~\cite{Freitas:2016sty}. Here we will illustrate the main points for the
example of the $W$ boson to discuss this issue. The propagator pole is defined
by
\begin{align}
Z_W(k^2-M^2_{W,0}) + \Sigma^W_T(k^2) = 0 
\qquad\text{for}\quad k^2 = M_W^2 - iM_W\Gamma_W \label{compm1}
\end{align}
The real part of the complex pole can be interpreted as the renormalized mass $M_W$, whereas the
imaginary part is associated with the decay width, $\Gamma_W$. Defining the OS
mass in this way ensures that it is well-defined and gauge-invariant to all
orders in perturbation theory, since the propagator pole is an analytic property
of the physical $S$-matrix
\cite{Willenbrock:1991hu,Sirlin:1991fd,Stuart:1991xk,Gambino:1999ai}.

What is the implication of the complex pole for the counterterms $\delta Z_W$
and $\delta M_W^2$? To answer this question, let us assume that $\Gamma_W \ll
M_W$\footnote{Numerically, $\Gamma_W/M_W \approx 2.5\%$ in the SM.}. Then one can
expand \eqref{compm1} as
\begin{align}
Z_W(M_W^2-iM_W\Gamma_W-M_W^2-\delta M_W^2) + \Sigma^W_T(M_W^2) -
 iM_W\Gamma_W \, \Sigma^{W\prime}_T(M_W^2) + {\cal O}(\Gamma_W^2) = 0
 \label{compm2}
\end{align}
where $\Sigma^{W\prime}_T(k^2) = \frac{\partial}{\partial(k^2)}\Sigma^W_T(k^2)$.
Taking the imaginary part of \eqref{compm2} one obtains
\begin{align}
&Z_W M_W \Gamma_W \approx \text{Im}\,\Sigma^W_T(M_W^2)
 - M_W\Gamma_W \; \text{Re}\,\Sigma^{W\prime}_T(M_W^2) \\
&\Rightarrow\quad \Gamma_W \approx \frac{\text{Im}\,\Sigma^W_T(M_W^2)}{M_W [Z_W +
\text{Re}\,\Sigma^{W\prime}_T(M_W^2)]} \label{compm3}
\intertext{$i.\,e.$ this provides a prescription for computing the total decay
width. On the other hand, the real part of \eqref{compm2} leads to}
&Z_W\,\delta M_W^2 \approx \text{Re}\,\Sigma^W_T(M_W^2)
 + M_W\Gamma_W \; \text{Im}\,\Sigma^{W\prime}_T(M_W^2) \label{compm4}
\end{align}
The last term in \eqref{compm4} would not be present for a stable particle.
However, for an unstable particle, its inclusion is important to ensure that the
renormalized mass is well-defined and gauge-invariant.

Eqs.~\eqref{compm3} and \eqref{compm4} depend on the field renormalization
counterterm $\delta Z_W$. However, it becomes ill-defined when taking into
account the width $\Gamma_W$, because we do not know whether we should demand
that it compensates the self-energy correction for $p^2=M_W^2$ or for
$p^2=M_W^2-iM_W\Gamma_W$. The latter may seem preferrable because it is 
the gauge-invariant pole of the propagator, but what are we to make of an
external particle with complex momentum?

The problem occurs because we have explicitly taken into account the fact that
the $W$ boson is unstable. But in this case it cannot be an asymptotic external
state, because it will decay rather rapidly! Instead, we should consider a
process where the production and decay of the $W$ boson is included, so that it
occurs only as an internal particle. An example would be $u\bar{d} \to W^+ \to
\mu^+\nu_\mu$. When computing this process, $\delta Z_W$ occurs in the
intermediate $W$ propagator, but also in the initial-state $u\bar{d}W$ vertex
and in the final-state $W\mu^+\nu_\mu$ vertex. Summing up all these
contributions, one can easily verify that $\delta Z_W$ drops out in the total
result, and we never need to provide an explicit expression for it.

\bigskip

\noindent
\begin{minipage}[b]{.7\textwidth}
\paragraph{Tadpole renormalization:} A large number of loop diagrams contains
so-called \emph{tadpoles}, which are sub-diagrams with one external leg. An
example for the process $\mu^- \to e^-\nu_\mu\bar{\nu}_e$ is shown to the right.
In a practical calculation, these diagrams constitute a large fraction of the
total number of diagrams and they signficantly increase the size of 
intermediate algebraic expressions.
\end{minipage}
\hfill\raisebox{1em}{%
\epsfig{figure=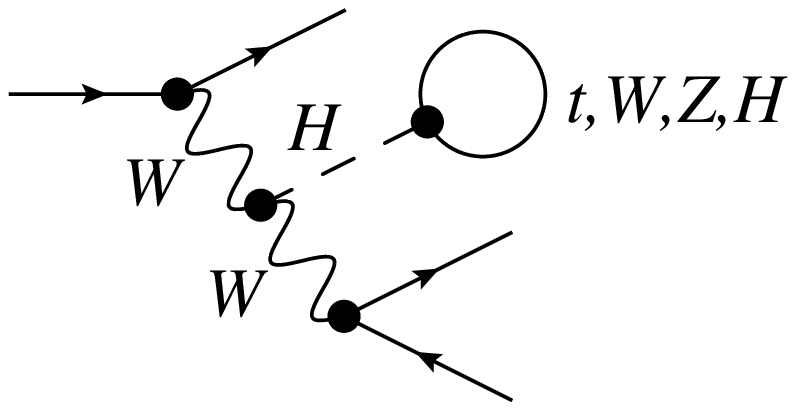, width=.25\textwidth}}

Fortunately, these tadpoles can be absorbed by renormalizing the Higgs vev, 
\begin{align}
v_0 = v + \delta v \label{vevren}
\end{align} 
Introducing this additional counterterm will break any
tree-level relationships between the vev and other parameters. For example, for
the bare Higgs potential,
\begin{align}
V_0 &= -\mu_0|\phi|^2 + \lambda_0|\phi|^4
\end{align}
one has $v_0 = \sqrt{\mu_0^2/\lambda_0}$, but this relationship between $v$,
$\lambda$ and $\mu$ can be modified at higher orders without causing any
problems since $v$ is not an observable, and its numerical value does not have
any direct physical meaning. Therefore, we are free to choose the counterterm
$\delta v$ at will during the computation of any physical observables, without
affecting the final result.

To see how this can be used to eliminate tadpole diagrams, let us write the
Higgs doublet field as
\begin{align}
\phi &= \begin{pmatrix} G^+ \\ \frac{1}{\sqrt{2}}(v+H+G^0) \end{pmatrix}
\end{align}
where $G^+,G^0$ are the Goldstone fields. Using $\mu_0^2 = \frac{1}{2}M_{H,0}^2
= \frac{1}{2}(M_H^2 + \delta\tilde{M}_H^2)$, and expanding the bare Higgs
potential to one-loop order, one finds
\begin{align}
V_0 &= V + \delta V + {\cal O}(\delta X^2), \\
V &= -\frac{M_H^2}{2}|\phi|^2 + \frac{M_H^2}{2v^2}|\phi|^4, \\
\delta V &= \text{const.} - \underbrace{M_H^2\,\delta v}_{\equiv\, \delta t}\,H
 +
 \underbrace{\Bigl(\frac{\delta\tilde{M}_H^2}{2}-\frac{3}{2}M_H^2\,\frac{\delta
 v}{v} \Bigr)}_{\equiv\, \delta M_H^2/2}H^2 - \frac{M_H^2\,\delta v}{2v}(G_0^2
 + 2G^+G^-) + \text{interact.} \label{delV}
\end{align} 
To avoid clutter, the constant term (which is physically irrelevant) and any
interaction terms involving three or more scalar fields have not been spelled
out in eq.~\eqref{delV}. The term linear in $H$ produces a tadpole-like
counterterm Feynman rule:
\begin{align}
& \raisebox{-1em}{\epsfig{figure=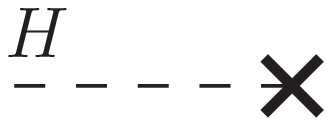, width=5em, 
 bb=245 662 360 712, clip=}} = i\,\delta t
\intertext{Now one can choose $\delta t$ (or, equivalently, $\delta v$) such
that the sum of the tadpole loop diagrams plus this counterterm vanishes:}
&\raisebox{-1em}{\epsfig{figure=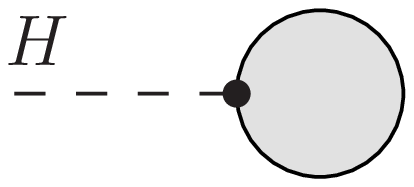, width=5em, 
 bb=245 662 360 712, clip=}} \;+
\raisebox{-1em}{\epsfig{figure=tadct.eps, width=5em, 
 bb=245 662 360 712, clip=}} \; =\; 0
 \end{align}
When adopting this convention, no tadpole diagrams need to be taken into account
in an actual calculation of a physics process.

As can be seen in \eqref{delV}, $\delta v$ also modifies the Higgs mass
counterterm, but this shift can be absorbed by redefining this mass counterterm.
The new counterterm $\delta M_H^2$ can be derived from the standard OS
renormalization condition without needing to worry about tadpoles.

However, $\delta v$ also generates a fictious mass for the Goldstone bosons
($G^0,G^\pm$), as also shown in \eqref{delV}. 
Since these fields are \emph{a priori} massless, this fictious
mass cannot be absorbed by any redefinition of other parameters. While this is
not explicitly shown in \eqref{delV}, additional non-trivial contributions
proportional to $\delta v$ also appear in self-interactions of the Goldstone
scalars. These Goldstone mass and vertex corrections are a (small) price to pay
for renormalizing away the tadpole diagrams. For a complete list of Feynman
rules modified by $\delta t$ ($\delta v$), see $e.\,g.$
Ref.~\cite{Denner:1991kt}.

\paragraph{Exercise:} Determine the contributions of $\delta t$ and $\delta
M_H^2$ to the scalar three- and four-point interactions  in $\delta V$.
\label{dV}

%%%%%%%%%%%%%%%%%%%%%%%%%%%%%%%%%%%%%%%%%%%%%%%%%%%%%%%%%%%%%%%%%%%%%%%%%%%%%%%

\subsection{Other Renormalization Schemes}
\label{otherren}

While the OS scheme has certain advantages, by relating renormalized SM
parameters to physical observables, a range of other renormalization schemes are
frequently adopted in the literature. They each come with specific advantage and
disadvantages (indicated by 
+\hspace{-2.1ex}$\bigcirc$ and 
$-$\hspace{-2.1ex}$\bigcirc$ below, respectively).

\paragraph{\msbar\ scheme:} All already mentioned on page~\pageref{msbar1}, all
counterterms in the \msbar\ scheme have the form
\begin{align}
\delta X &= (4\pi e^{-\gamma_{\rm E}})^{L\varepsilon}
\Bigl(\frac{C_L}{\varepsilon^L} + \frac{C_{L-1}}{\varepsilon^{L-1}} +
... + \frac{C_1}{\varepsilon} \Bigr),
&& \varepsilon = \frac{2}{4-d}, \quad L = \text{loop order}
\end{align}
$i.\,e.$ they simply subtract the divergent pieces of an amplitude but do not
contain any non-trivial finite terms. The factor with $4\pi$ and $\gamma_{\rm
E}$ in front is included to cancel similar terms that universally appear for any
divergent loop integral in dimensional regularization.
\begin{itemize}
\item[+\hspace{-1.7ex}$\bigcirc$]
The dependence on the scale $\mu$ of dimensional regularization, $d^4k \to
\mu^{4-d} d^dk$, is not cancelled by the counterterms, so that the \msbar\
couplings and masses depend on the choice of $\mu$. The $\mu$-dependence can be
described by the renormalization group, which allows one to resum dominant terms
in some calculations and to study phase transitions.
\item[+\hspace{-1.7ex}$\bigcirc$]
In some cases, when renormalizing the relevant parameters of a physical process
in the \msbar\ scheme, the perturbation series for this process converges better
than in the OS scheme.
\item[$-$\hspace{-1.7ex}$\bigcirc$]
One needs an additional calculation to relate a \msbar\ parameter to an oservable,
in order to determine the numerical value for this parameter from experiment.
\end{itemize}

%------------------------------------------------------------------------------
\begin{figure}
\centering
\begin{tabular}{c@{\hspace{3em}}c@{\hspace{3em}}c@{\hspace{3em}}c}
\epsfig{figure=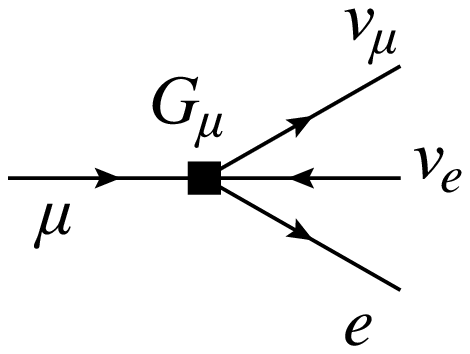, width=1in} &
\epsfig{figure=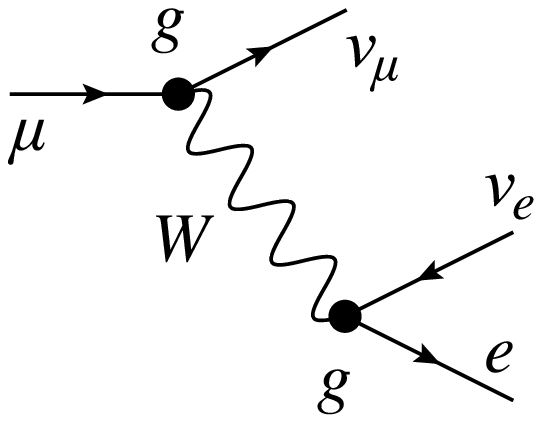, width=1.2in} &
\epsfig{figure=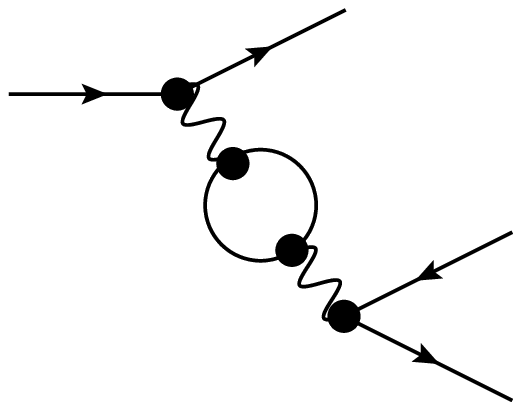, width=1.2in} &
\epsfig{figure=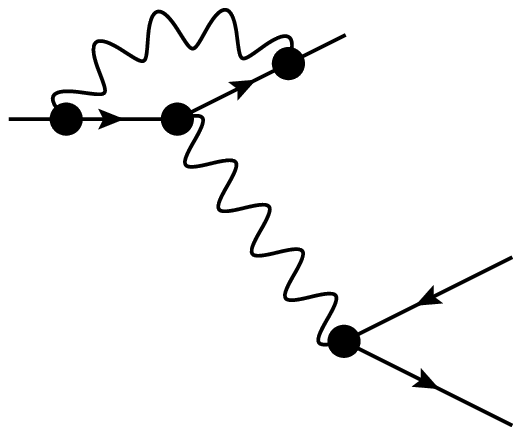, width=1.2in} \\
(a) & (b) & (c) & (d) 
\end{tabular}
\caption{Diagrams for muon decay (a) in the Fermi model, (b) at tree-level in
the SM, and (c,d) contributing to the one-loop corrections in the SM.}
\label{mudec}
\end{figure}
%------------------------------------------------------------------------------
\paragraph{\boldmath $G_\mu$ scheme:} The Fermi constant $G_\mu$ describes the
decay of muons as an effective four-fermion interaction described by the
Lagrangian
\begin{align}
{\cal L}_{\rm Fermi} = \frac{G_\mu}{2\sqrt{2}}
 \bigl(\overline{\psi}_{\nu_\mu}\gamma_\lambda\omega_-\psi_\mu\bigr)
 \bigl(\overline{\psi}_e\gamma_\lambda\omega_-\psi_{\nu_e}\bigr)
\end{align}
where $\omega_- \equiv (1-\gamma_5)/2$. In Feynman diagrammatic form, the muon
decay process generated by this interaction in shown in Fig.~\ref{mudec}~(a).
In the SM, the four-fermion interaction
is instead mediated by $W$-boson exchange at tree-level, see Fig.~\ref{mudec}~(b).
Therefore, $G_\mu$ can be expressed in terms of SM parameters,
\begin{align}
\frac{G_\mu}{\sqrt{2}} = \frac{g^2}{8M_W^2}(1+\Delta r) \label{GmuSM}
\end{align}
where $\Delta r$ accounts for corrections beyond tree-level (see
Fig.~\ref{mudec}~(c,d) for example diagrams).
This relation can be used to express the weak coupling $g$ in terms of $G_\mu$:
\begin{align}
g^2 = 4\sqrt{2}\,g_\mu M_W^2 (1+\Delta)^{-1} \label{gtoGmu}
\end{align}
When computing electroweak radiative corrections, instead of writing them as a
series in powers of $g=e/\sw$, one can employ \eqref{gtoGmu} to represent them
as a series in powers of $G_\mu$.
\begin{itemize}
\item[+\hspace{-1.7ex}$\bigcirc$] $G_\mu = 1.1663787(6)\times
10^{-5}\,\text{GeV}^{-2}$ is precisely known from measurement
\cite{Zyla:2020zbs}.
\item[+\hspace{-1.7ex}$\bigcirc$] The leading corrections in $\Delta r$ may
(partially) cancel similar terms in other observables. For example, when
computing the $W$ decay rate, the one-loop corrections are much smaller in the
$G_\mu$ scheme than in the OS scheme of the previous subsection \cite{Denner:1991kt}.
\item[$-$\hspace{-1.7ex}$\bigcirc$] One needs to include the corrections to muon
decay ($\Delta r$) in the calculation of any other observable.
\end{itemize}

%%%%%%%%%%%%%%%%%%%%%%%%%%%%%%%%%%%%%%%%%%%%%%%%%%%%%%%%%%%%%%%%%%%%%%%%%%%%%%%

\section{Electroweak Precision Observables}
\label{ewpos}

The term \emph{electroweak precision observable} (EWPO) refers to a set of
quantities that have been measured with high precision (typically at the
per-mille level or better) and that are related to properties of the electroweak
($W$ and $Z$) gauge bosons. In general, they also include a number of quantities
that are not stricty instrinsic to the electroweak sector, but that are needed
to make predictions for EWPOs within the SM. These are often called ``input
parameters.''

Another rationale for distinguishing between input parameters and ``genuine''
EWPOs is the expectation that input parameters are unlikely to be significantly
affected by new physics beyond the SM (possible reasons include: their
measurement is based on kinematical features; they are protected by symmetries;
new physics decouples due to effective field theory arguments). On the other
hand, the genuine EWPOs can get modified by a large range of beyond-the-SM (BSM)
models. In fact, one of the main motiviations for studying EWPOs is their
potential to constrain new physics by comparing measurement data with
theoretical SM predictions.

In the following subsection, the relevant input parameters will be discussed one
by one, before giving an overview of the most important genuine EWPOs in the
remainder of this chapter.

%%%%%%%%%%%%%%%%%%%%%%%%%%%%%%%%%%%%%%%%%%%%%%%%%%%%%%%%%%%%%%%%%%%%%%%%%%%%%%%

\subsection{Input Parameters}

A typical choice of input parameters for electroweak precision studies is:
$\alpha = \frac{e^2}{4\pi}$, $G_\mu$, $\alpha_{\rm s}$, $M_Z$, $M_H$, $m_t$.
The masses of any fermions besides the top quark ($m_f,\,f\neq t$) are generally
negligible in electroweak physics since their impact is suppressed by powers of
$m_f^2/M_W^2$, with the exception of $\Delta\alpha$, where they contribute
logarithmically, see page~\pageref{deltaalpha}.

\paragraph{Fine structure constant \boldmath $\alpha$:}
There are two leading methods for determining the value of $\alpha$:
\begin{itemize}
\item From the electron magnetic moment $a_e = \frac{g_e-2}{2}$
\cite{Aoyama:2019ryr}, which has been
theoretically computed to very high precision:
\begin{align}
a_e = \frac{\alpha}{2\pi} + A_2\alpha^2 + A_3\alpha^3 + A_4\alpha^4 +
A_5\alpha^5 + C_{\rm EW}\frac{m_e^2}{M_W^2} + C_{\rm
had}\frac{m_e^2}{\Lambda_{\rm QCD}^2} ... \label{aeth}
\end{align}
The coefficients $A_i$ denote $i$-loop QED loop corrections, which have been
computed to five-loop order \cite{Aoyama:2012wk,Aoyama:2012qma,Baikov:2013ula}. Electroweak corrections, denoted by the
term with $C_{\rm EW}$, are suppressed by the small electron mass. Similarly,
hadronic corrections, denoted by the
term with $C_{\rm had}$, first enter at the two-loop level, and they are
suppressed by the ratio $m_e^2/\Lambda_{\rm QCD}^2$, with $\Lambda_{\rm QCD}
\sim {\cal O}(1\,\text{GeV})$. Both of these contributions are negligible at
current levels of precision.

Comparing \eqref{aeth} to precise measurements of $g_e$ using Penning traps
\cite{Hanneke:2008tm}, one finds \cite{Mohr:2015ccw,Aoyama:2017uqe}
\begin{align}
\alpha^{-1} = 137.035\,999\,174(35) \label{alpha1}
\end{align}
where the numbers in brackets indicate the uncertainty in the last quoted
digits.
\item An independent determination of $\alpha$ can be obtained from the defining
formula for the Rydberg constant, $R_\infty$:
\begin{align}
\alpha^2 = \frac{R_\infty}{2c}\,\frac{m_{\rm At}}{m_e}\,\frac{h}{m_{\rm At}}
\end{align}
Here $m_{\rm At}$ is the mass of some atom.
The fine structure constant can be determined by using precise values for
$R_\infty$ (from atomic spectroscopy), $m_{\rm At}$ and $m_e$ (in atomic units),
and $h/m_{\rm At}$ (from atom interferometry). The limiting factor, in terms of
precision, is the measurement of $h/m_{\rm At}$, which recently has been
significantly improved for Cs-133 atoms \cite{Parker:2018vye}, resulting in
\begin{align}
\alpha^{-1} = 137.035\,999\,046(27) \label{alpha2}
\end{align}
\end{itemize}
The two values \eqref{alpha1} and \eqref{alpha2} exhibit a 2.5$\sigma$ tension.

\begin{figure}
\centering
\begin{tabular}{c@{\hspace{3em}}c@{\hspace{3em}}c@{\hspace{3em}}c}
\raisebox{-.1in}{\epsfig{figure=mudec_Gmu.eps, width=1in}} &
\epsfig{figure=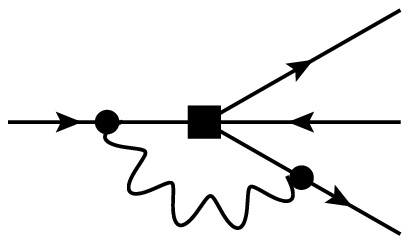, width=1in} &
\raisebox{-.05in}{\epsfig{figure=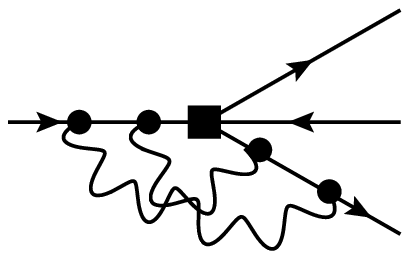, width=1in}} \\
(a) & (b) & (c)
\end{tabular}
\caption{Muon decay in the Fermi model: (a) leading order diagram, and sample
diagrams for the (b) one-loop and (c) two-loop QED corrections.}
\label{mudec2}
\end{figure}

\paragraph{Fermi constant \boldmath $G_\mu$:} 
As already mentioned in section~\ref{otherren}, the Fermi constant gives the
strength of an effective four-fermion interaction, which can be
extracted from muon decay. Besides the leading-order diagram in
Fig.~\ref{mudec2}~(a), there are also significant QED corrections as illustrated
in Fig.~\ref{mudec2}~(b,c). The corrections are known to next-to-next-to-leading
(NNLO) order \cite{vanRitbergen:1999fi,Steinhauser:1999bx,Pak:2008qt}. Combining
these with the measured value of the muon lifetime, $\tau_\mu$,
\cite{Webber:2010zf}, one obtains 
\begin{align}
G_\mu = 1.1663787(6) \times 10^{-5} \mbox{ GeV}^{-2}
\end{align}
where the uncertainty is dominant by the experimental measurement of $\tau_\mu$,
whereas the estimated theory error from missing higher orders is sub-dominant.
%------------------------------------------------------------------------------
%------------------------------------------------------------------------------

\paragraph{Strong coupling \boldmath $\alpha_{\rm s}=g_{\rm s}^2/(4\pi)$:} There
are many independent methods for its determination. For a complete review, see
chapter 9 of Ref.~\cite{Zyla:2020zbs}. In the following, a few of the most
precise methods are listed:
\begin{itemize} 
\item The currently most precise
approach uses lattice QCD calculations. Two recent studies yield
\begin{align}
\text{Lattice:} \qquad \alpha_{\rm s} &= 0.1185 \pm 0.0008 \quad \text{\cite{Bruno:2017gxd}} \\
 \alpha_{\rm s} &= 0.1172 \pm 0.0011 \quad \text{\cite{Zafeiropoulos:2019flq}}
\end{align}
\item Differential distributions (event shapes) of $e^+e^- \to \text{jets}$ and
deep inelastic scattering (DIS), using NNLO QCD corrections. These approaches yield
values of $\alpha_{\rm s} \approx 0.114$ on average, significantly below the
numbers obtained with other methods. Possible issues include large
non-perturbative QCD uncertainties (for $e^+e^- \to \text{jets}$) and scheme
dependence and parametrization of parton distribution functions (for DIS). See
chapter 9 of Ref.~\cite{Zyla:2020zbs} for more details.
\item From the branching fraction of taus into hadrons one obtains
\begin{align}
\tau\text{ decays:} \qquad \alpha_{\rm s} &= 0.117 \pm 0.002
\end{align} 
(see section 10 of Ref.~\cite{Zyla:2020zbs}). This determination is subject
to non-perturbative hadronic uncertainties, $e.\,g.$ from violations of
quark-hadron duality.
\item EWPOs, in particular the branching ratio $R_\ell \equiv
\Gamma[Z\to\text{had.}]/\Gamma[Z\to\ell^+\ell^-]$ ($\ell=e,\mu,\tau$), yield
\begin{align}
\text{Electroweak precision:} \qquad \alpha_{\rm s} &= 0.1221 \pm 0.0027
\end{align}
(see section 10 of
Ref.~\cite{Zyla:2020zbs}). This method has negligible QCD uncertainties
(both perturbative and non-perturbative), but since $R_\ell$ is a high-energy
observable, it is more likely to be impacted by new physics beyond the SM.
\end{itemize}

\paragraph{Top-quark mass \boldmath $m_t$:} The currently most precise
measurements are based on the invariant mass distribution ($m_{\rm inv}$) of the
top decay products at LHC (for a review see the section ``Top Quark'' in
Ref.~\cite{Zyla:2020zbs}). This approach yields a result that is
numerically very close to the OS (pole) mass \cite{Hoang:2018zrp}. However, the
OS mass of the top quark is theoretically not well-defined due to the presence
of non-perturbative QCD contributions to the top-quark self-energy in
\eqref{dysonf3}. These effects, called \emph{renormalons}, are typically of the
order of ${\cal O}(\Lambda_{\rm QCD}) \sim 300$~MeV, where $\Lambda_{\rm QCD}$
is the scale where $\alpha_{\rm s}$ becomes non-perturbative
\cite{Beneke:2016cbu}. Therefore, when trying to use the peak of the $m_{\rm
inv}$ distribution as an input for other calculations, there necessarily is an
ambiguity of the same order.

\begin{figure}
\centering
\epsfig{figure=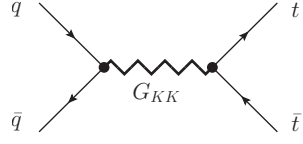, width=1.5in, bb=190 615 404 720, clip=true}
\caption{Example of a new physics contribution to $t\bar{t}$ production at the
LHC, due to a Kaluza-Klein excitation of the gluon.}
\label{gkk}
\end{figure}

The problem of the top-quark mass definition can be circumvented by measuring a
more inclusive observable, such as the total $t\bar{t}$ cross-section,
$\sigma_{t\bar{t}}$, at the LHC. $\sigma_{t\bar{t}}$ can be described in terms of
the \msbar\ mass $m_t^{\msbar}$, which is free of the renormalon ambiguity.
However, it may be affected by possible new physics effects, such as heavy new
physics particles in the $s$-channel, which are predicted by theories
with extra dimensions and other models (see Fig.~\ref{gkk}).
%------------------------------------------------------------------------------
%------------------------------------------------------------------------------

%------------------------------------------------------------------------------
\begin{figure}
\centering
\begin{tabular}{c@{\hspace{.5in}}c}
\epsfig{figure=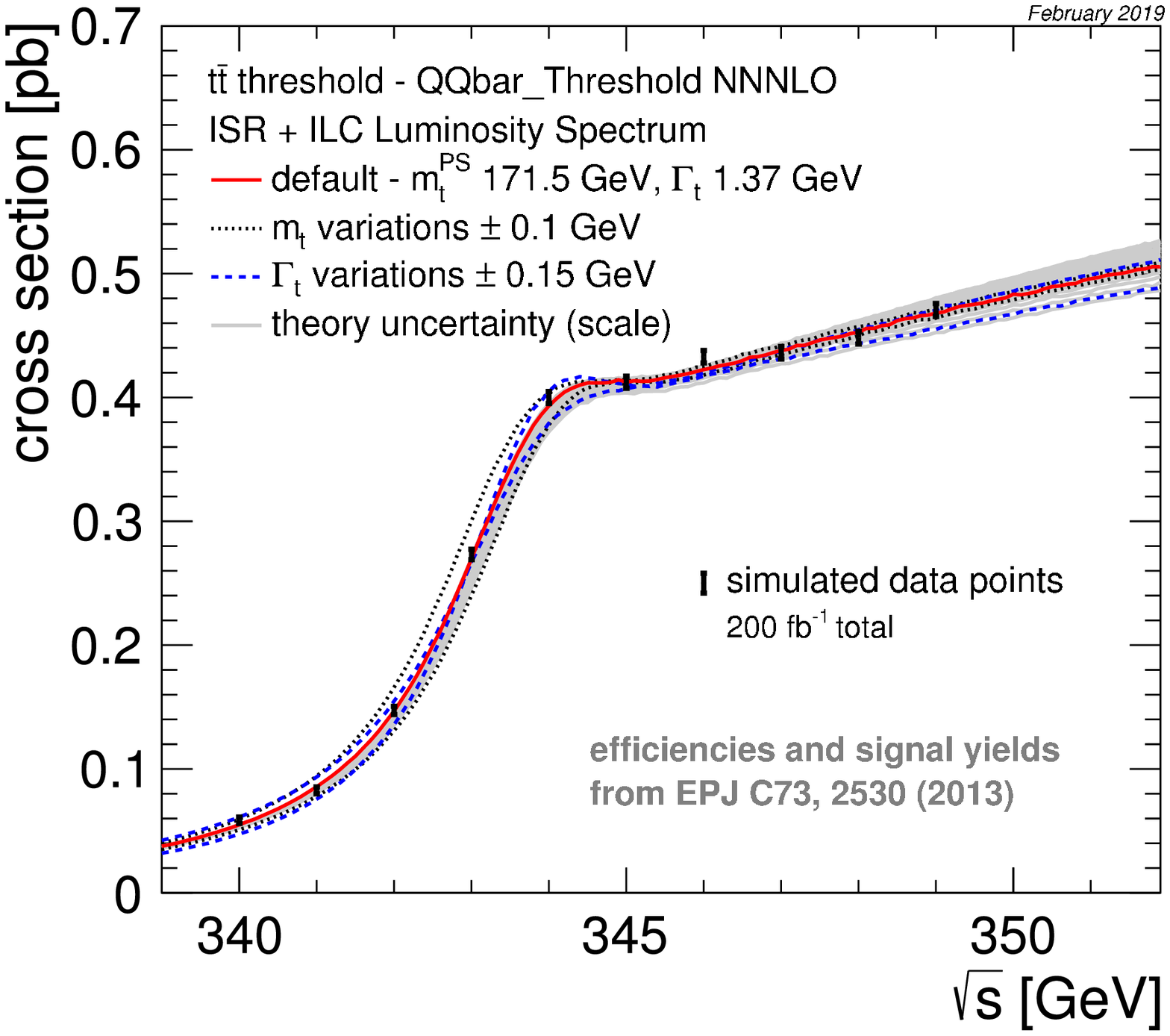, width=8cm, bb=255 95 822 575} &
\raisebox{1in}{\epsfig{figure=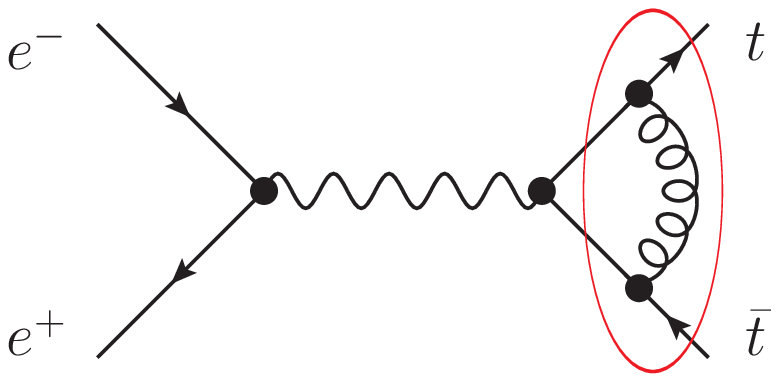, width=1.8in, 
 bb=180 610 404 720, clip=true}} \\[-1ex]
 (a) & (b)
\end{tabular}
\caption{(a) Illustration of a 10-point threshold scan for $e^+e^- \to t\bar{t}$ at ILC
(figure taken from Ref.~\cite{Simon:2019axh}). (b) Sample diagram of a
gluon-exchange contribution to the $t\bar{t}$ bound-state effect near threshold.}
\label{eett}
\end{figure}
%------------------------------------------------------------------------------
At future $e^+e^-$ colliders with a center-of-mass energy of at least 350~GeV, a
precise, well-defined measurement of $m_t$ can be performed, that is largely
unaffected by BSM physics. This is achieved through a \emph{threshold scan},
measuring $\sigma_{t\bar{t}}$ at different values of the center-of-mass energy
$\sqrt{s}$, see Fig.~\ref{eett}~(a). The shape of $\sigma_{t\bar{t}}$ as a function of $\sqrt{s}$ can be
predicted with high precision in terms of $m_t^{\msbar}$, including NNNLO QCD as
well as NLO and leading NNLO electroweak corrections
\cite{Beneke:2015kwa,Beneke:2016kkb}. The small bump in the lineshape at the
threshold is caused by 1S bound-state effects from gluon exchange\footnote{Here
the spectroscopic notation ``1S'' is used for the lowest-energy mode with zero
orbital angular momentum.}, as
illustrated in Fig.~\ref{eett}~(b). Note that this is not a true bound state
since the decay width of the top quark is larger than the binding energy, but it
still leads to a an enhancement of the cross-section at the would-be bound-state
energy.

\paragraph{Z-boson mass \boldmath $M_Z$:} The most precise determination of
$M_Z$ is obtained from measurements of the cross-section for $e^+e^- \to
f\bar{f}$ at different center-of-mass energies at LEP.

Ignoring $\gamma$--$Z$ mixing for the moment, this process can be described by
the generic Feynman diagram below, yielding
\begin{align}
&\raisebox{-0.25in}{\epsfig{figure=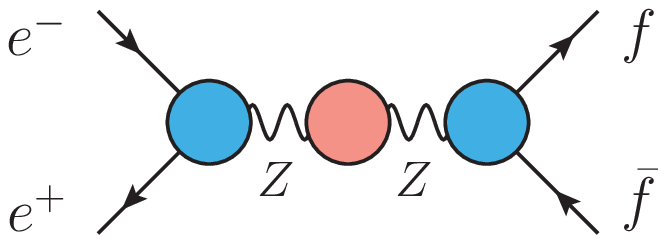, width=1.5in, 
 bb=200 648 390 720, clip=true}}
& 
\sigma_f(s) &= \biggl|\frac{\color{blue} R(s)}{s-M_Z^2-\color{red} \delta M_Z^2 +
\Sigma^Z_T(s)/Z_{ZZ}}\biggr|^2 \label{eezff1} \\[-1ex]
&&&= \biggl|\frac{R(M_Z^2)}{s-M_Z^2+iM_Z\Gamma_Z} + \text{non-res.}\biggr|^2 \label{eezff2} 
\end{align}
In line \eqref{eezff1}, the red blob and terms indicates the contribution from
the Z self-energy (including counterterms), whereas the blue blobs and term
denotes contributions from vertex corrections. The amplitude inside the modulus
brackets $||$ in \eqref{eezff1} has a complex pole (see
page~\pageref{unstable}). Expanding about this pole and using $\Gamma_Z \ll M_Z$
yields the expression in
\eqref{eezff2}, which has a resonant piece and an infinite series of terms
suppressed by powers of $(s-M_Z^2)$ and $\Gamma_Z/M_Z$, which are not explicitly spelled out here.

Including $\gamma$--$Z$ mixing requires the replacement of $\Sigma_T^Z(s)$ with
\label{mz}
\begin{align}
\Sigma_T^Z(s) &\to
 \Sigma_T^Z(s) - \frac{[\hat\Sigma_T^{AZ}(s)]^2}{s+\hat\Sigma_T^A(s)}\,, \\
\hat\Sigma_T^{AZ}(s) &= \Sigma_T^{AZ}(s) + \tfrac{1}{2}
 \delta Z_{ZA}\sqrt{Z_{ZZ}}(s-M_Z^2-\delta M_Z^2) + s\tfrac{1}{2}
 \delta Z_{AZ}\sqrt{Z_{AA}}\,, \\
\hat\Sigma_T^A(s) &= \Sigma_T^A(s) + s\tfrac{1}{2} \delta Z_{AA}
 + \tfrac{1}{4}(\delta Z_{ZA})^2(s-M_Z^2-\delta M_Z^2)
\end{align}
Even though the expressions for the counterterms are rather lengthy, the result
still takes the general form in \eqref{eezff2}, since this result only relies on
the presence of a complex pole in the amplitude. 

\begin{figure}
\centering
\epsfig{figure=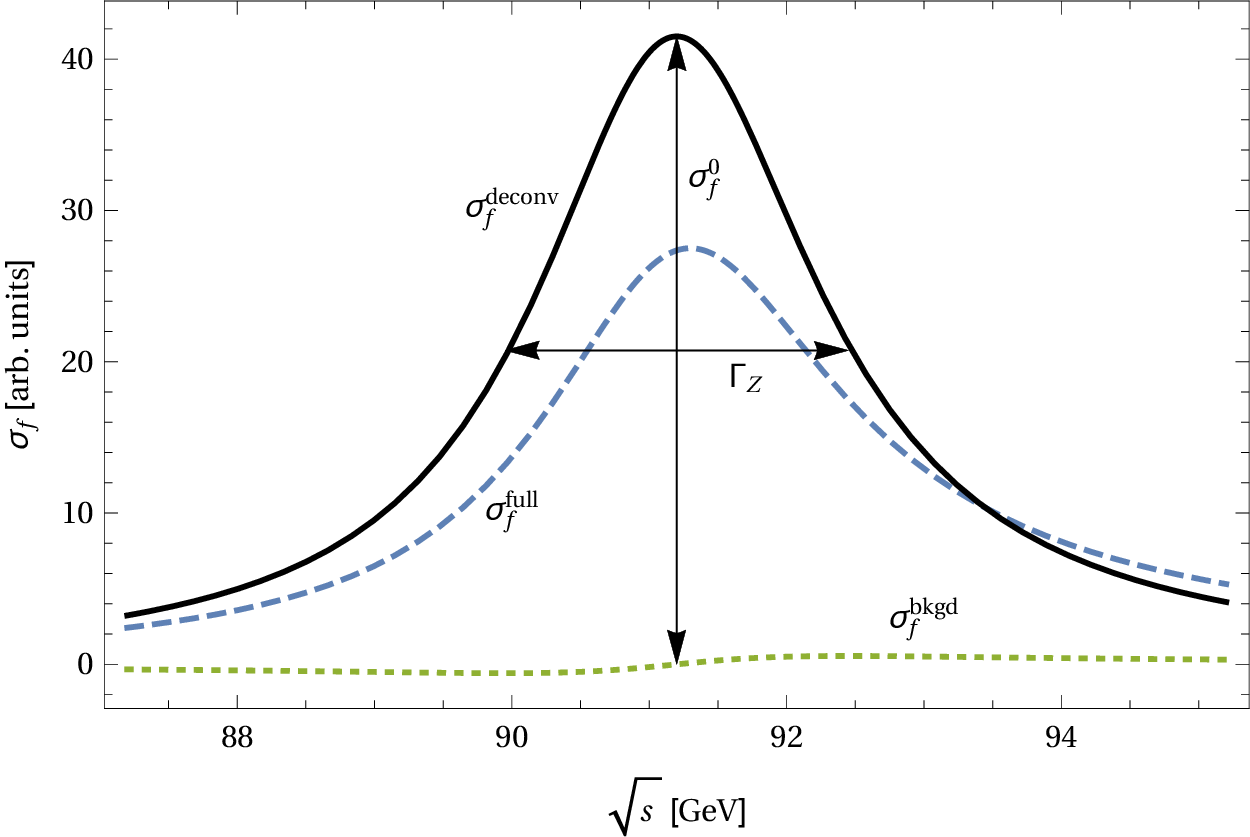, width=12cm}
\vspace{-1ex}
\caption{Illustration of $Z$-pole cross-section line-shape (not to scale). The
solid (dashed) line indicates the line-shape without (with) initial-state QED
radiation. The dotted line depicts backgrounds from photon exchange and box
contributions to the cross-section (without initial-state radiation). [Figure
taken from Ref.~\cite{Freitas:2016sty}.]
\label{linez}} 
\end{figure}

Carrying out the square in \eqref{eezff2} yields
\begin{align}
\sigma_f(s) &= \frac{|R(M_Z^2)|^2}{(s-M_Z^2)^2 + M_Z^2\Gamma_Z^2} +
\text{non-res.} \label{bw1}
\end{align}
which is called a \emph{Breit-Wigner} resonance, see the solid curve in
Fig.~\ref{linez}.
%-------------------------------------------------------------
%-------------------------------------------------------------

By fitting this curve to experimental measurements of $\sigma_f$ at three or
more values of $\sqrt{s}$, one can determine $M_Z$ and $\Gamma_Z$ at high
precision. However, in the experimental studies at LEP, a different
parametrization of the Breit-Wigner resonance has been used,
\begin{align}
\sigma_f(s) &= \frac{R'^2}{(s-m_Z^2)^2 + s^2\gamma_Z^2/m_Z^2} + 
\text{const.} \label{bw2}
\end{align}
with the results \cite{Zyla:2020zbs,Janot:2019oyi}
\begin{align}
m_Z &= 91.1876\pm 0.0021 \,\text{GeV}, &
\gamma_Z &= 2.4942\pm 0.0023 \,\text{GeV} \label{mzexp}
\end{align}
When ignoring the non-resonant terms, the two forms \eqref{bw1} and \eqref{bw2}
are fully equivalent, but the mass and width parameters are different. The
relation is given by \cite{Bardin:1988xt}
\begin{align}
\begin{aligned}
M_Z &= m_Z(1+\gamma_Z^2/m_Z^2)^{-1/2} \approx m_Z - 34\,\text{MeV}, \\
\Gamma_Z &= \gamma_Z(1+\gamma_Z^2/m_Z^2)^{-1/2} \approx m_Z - 0.9\,\text{MeV} \\
\end{aligned} \label{mztrans}
\end{align}
Thus, whenever aiming to use \eqref{mzexp} as inputs to a theory calculation,
one first needs to apply the translation \eqref{mztrans}\footnote{The same is true
for W mass measurements at colliders.}.

\paragraph{Exercise:} For each the quantities listed above, which of the following
concepts limits the influence of new physics in their determination: measurement
is based on kinematical features; protected by symmetries; new physics decouples
due to effective field theory arguments (see appendix for answer).
\label{input}

%%%%%%%%%%%%%%%%%%%%%%%%%%%%%%%%%%%%%%%%%%%%%%%%%%%%%%%%%%%%%%%%%%%%%%%%%%%%%%%

\subsection{Z-pole EWPOs}

Electroweak precision observables at the $Z$-pole are related to the vector and
axial-vector couplings of the $Z$--fermion interactions. For massless fermions,
these interactions have the form
\begin{align}
\overline{\psi}_f\,i\gamma_\mu(v_f - a_f \gamma_5) \, \psi_f
\end{align}
where the subscript $f$ labels the fermion type ($f=e,\mu,\tau,...$). At leading
order (Born level),
\begin{align}
v_f = e\frac{I^3_f-2\sw^2Q_f}{2\sw\cw}, \qquad
a_f e\frac{I^3_f}{2\sw\cw} \label{zcpl}
\end{align}
Here $Q_f$ is the fermion charge in units of the positron charge $e$, whereas
$I^3_f$ is the third component of weak isospin. Beyond Born level, $v_f$ and
$a_f$ receive radiative corrections within the SM and potentially also from BSM
contributions. Therefore, one can use precision measurements of these quantities
to constrain and potentially discover various typoes of new physics.

The following observables are useful to extract information about $v_f$ and
$a_f$ from data:
\begin{itemize}
\item \begin{minipage}[t]{.7\columnwidth}
The total $Z$ decay width, $\Gamma_Z$. According to \eqref{compm3},
$\Gamma_Z \propto \text{Im}\,\Sigma^W_T(M_W^2)$. Using the optical theorem,
which diagrammatically corresponds to cutting the self-energy diagram shown to
the right, the width is related to the matrix elements for the process $Z \to
f\bar{f}$, so that
\end{minipage}
\hfill
\raisebox{-2.5em}{\epsfig{figure=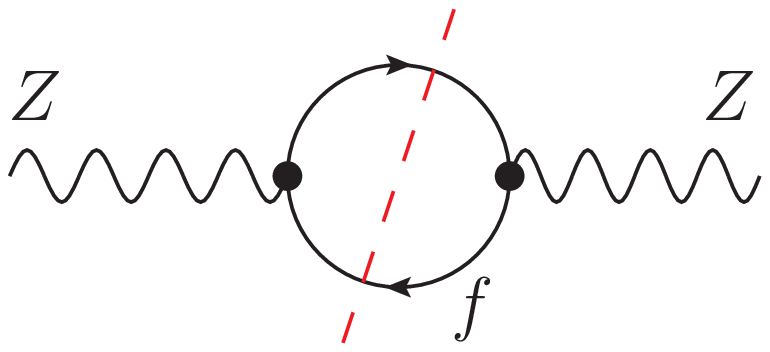, height=3.5em, bb=190 620 410 715,
 clip=true}}
\begin{align}
\Gamma_Z \propto \sum_f \bigl|{\cal M}[Z \to f\bar{f}]\bigr|^2
 = \sum_f \bigl( |v_f|^2 + |a_f|^2 \bigr)
\end{align}
\item The cross-section for $e^+e^- \to Z \to f\bar{f}$, which, up to a simple
phase-space and flux factor, can be written as
\begin{align}
\sigma_f(s) \propto \bigl( |v_e|^2 + |a_e|^2 \bigr)\,
\biggl|\frac{1}{s-M_Z^2+iM_Z\Gamma_Z+\Sigma^Z_T(s)+...}\biggr|^2
\bigl( |v_f|^2 + |a_f|^2 \bigr)
\end{align}
where the dots in the denominator refer to the $\gamma$--$Z$ mixing and
counterterms described on page~\pageref{mz}.
Near the $Z$ pole ($s \approx M_Z^2$) this expression can be recast into the
form
\begin{align}
\sigma_f(s) \approx 12\pi \frac{\Gamma_e\Gamma_f}{(s-M_Z^2)^2 + M_Z^2\Gamma_Z^2}
\equiv \sigma_f^0
\end{align}
where $\Gamma_f$ is the \emph{partial width} for the decay $Z \to f\bar{f}$ into
a particular fermion type $f$.
\item With polarized electron beams, one can measure the cross-section
separately for left- and right-handed polarized electrons:
\begin{align}
\sigma_{\rm L} \equiv \sigma[e^+e^-_{\rm L} \to f\bar{f}]
 \propto |v_e+a_e|^2\,
\biggl|\frac{1}{s-M_Z^2+iM_Z\Gamma_Z}\biggr|^2 \bigl( |v_f|^2 + |a_f|^2 \bigr)
 \\
\sigma_{\rm R} \equiv \sigma[e^+e^-_{\rm R} \to f\bar{f}]
 \propto |v_e-a_e|^2\,
\biggl|\frac{1}{s-M_Z^2+iM_Z\Gamma_Z}\biggr|^2 \bigl( |v_f|^2 + |a_f|^2 \bigr)
\end{align}
From this one can form a \emph{left-right asymmetry} where several important
systematic uncertainties, such as the lumonisity uncertainty or the detector
acceptance, cancel:
\begin{align}
A_{\rm LR} = \frac{\sigma_{\rm L}-\sigma_{\rm R}}{\sigma_{\rm L}+\sigma_{\rm R}}
 = \frac{2\,\text{Re}\{v_ea_e^*\}}{|v_e|^2 + |a_e|^2}
 = \frac{2\,\text{Re}\{v_e/a_e\}}{1+|v_e/a_e|^2} \equiv {\cal A}_e
\end{align}
Thus, in contrast to the decay width or the total cross-section, this asymmetry
yields information about the ratio of the vector and axial-vector couplings.
\begin{align}
&\text{At Born level:} & \frac{v_f}{a_f} &= 1-4|Q_f|\sw^2
\quad \Bigl[ = 1-4\sw^2 \text{ for } f=e\Bigr] \\
&\text{With higher orders:} & \text{Re}\,\frac{v_f}{a_f} &\equiv
 1-4|Q_f|\seff{f} \label{sweff}
\end{align}
where we have defined the \emph{effective weak mixing angle} $\seff{f}$ as the radiative
corrected verion of the on-shell weak mixing angle $\sw^2$.
\item
Without polarized beams, one can use the differential cross-section to obtain
information about the ratio $v_f/a_f$:
\begin{align}
\frac{d\sigma}{d\cos\theta} \propto \bigl( |v_e|^2 + |a_e|^2 \bigr)
 \bigl( |v_f|^2 + |a_f|^2 \bigr)(1+\cos^2\theta) + 4\,\text{Re}\{v_ea_e^*\}
 \,\text{Re}\{v_fa_f^*\}\cos\theta
\end{align}
where we have spelled out only the terms that depend
on the scattering angle $\theta$ (the angle between the momenta of the incident
$e^-$ and the outgoing $f$), whereas all other terms (such as the $Z$
propagator) subsumed in the unspecified proportionality factor.

The range of possible scattering angle can be divided into a forward and
backward hemisphere,
\begin{align}
&\sigma_{\rm F} \equiv = \int_0^1 d\cos\theta\;\frac{d\sigma}{d\cos\theta},
&
&\sigma_{\rm B} \equiv = \int_{-1}^0 d\cos\theta\;\frac{d\sigma}{d\cos\theta}
\end{align}
which then allows us to define the \emph{forward-backward asymmetry}
\begin{align}
A_{\rm FB}^f = \frac{\sigma_{\rm F}-\sigma_{\rm B}}{\sigma_{\rm F}+\sigma_{\rm B}}
 = \frac{3\,\text{Re}\{v_ea_e^*\}\,\text{Re}\{v_fa_f^*\}}{(|v_e|^2 + |a_e|^2)
    (|v_f|^2 + |a_f|^2}
 = \frac{3}{4} {\cal A}_e{\cal A}_f
\end{align}
\end{itemize}
The quantities introduced above ($\Gamma_Z$, $\Gamma_f$, $\sigma^0_f$, $A_{\rm
FB}^f$, $A_{\rm LR}$) are so-called \emph{pseudo-observables}. The reason for
this terminology is due to the fact that \emph{real observables} involve extra
effects:

\bigskip
\noindent
\begin{minipage}[b]{.6\textwidth}
\paragraph{Initial-state radiation (ISR):} There are corrections due to emission
of real and virtual photons off the incoming electron and photon. Photons that
are soft or collinear to one of the incoming particles lead to contributions
that are enhanced by terms involving logarithms of the form
\begin{align}
&\frac{2\alpha}{\pi}L \equiv \frac{2\alpha}{\pi}\ln\frac{s}{m_e^2} \approx 11\% 
&&\text{[for } s = M_Z^2]
\end{align}
\end{minipage}
\hfill\raisebox{2em}{%
\epsfig{figure=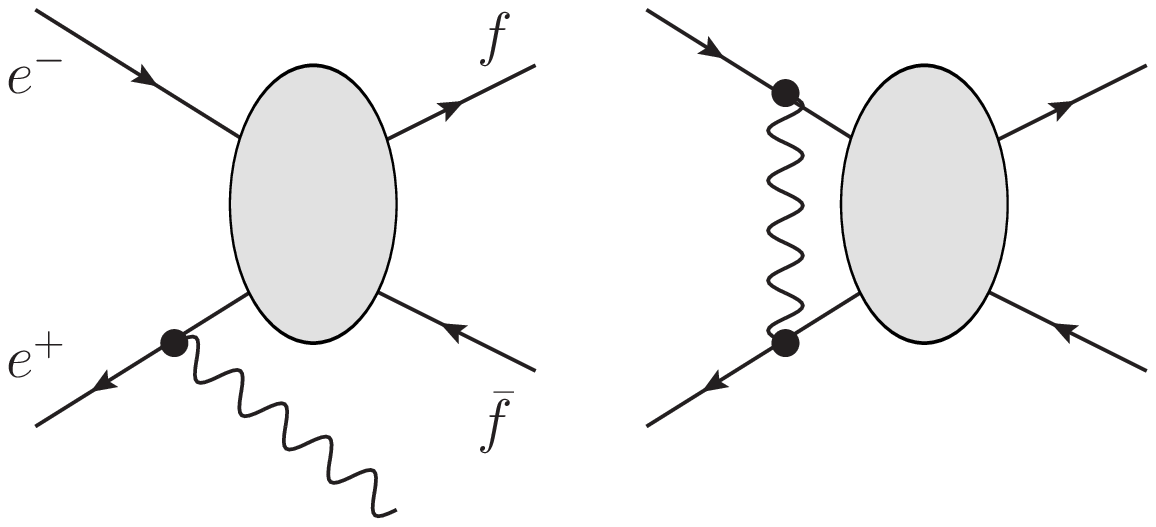, width=.35\textwidth, bb=130 560 465 715, clip=}}

\vspace{1ex}\noindent
The ISR effects can be taken into account through a convolution
\begin{align}
\sigma^{\rm full}_f(s) = \int_0^{1-4m_f^2/s} dx \; H(x) \, \sigma^{\rm
deconv}_f\bigl(s(1-x)\bigr) \label{isr}
\end{align}
The deconvoluted cross-section, $\sigma^{\rm deconv}_f$, is illustrated by the gray
blows in the diagrams above. The radiator function $H(x)$ contains the soft and
collinear photon contributions. It has the general form
\begin{align}
H(x) = \sum_n \Bigl(\frac{\alpha}{\pi}\Bigr)^n \sum_{m=0}^n h_{nm} (2L)^m
\end{align}
The leading logarithms (for $m=n$) are universal ($i.\,e.$ independent of the
specific process) and known to $n=6$ (see Ref.~\cite{Ablinger:2020qvo} and
references therein). Also some sub-leading terms are known for $e^+e^- \to
\gamma^*/Z* \to f\bar{f}$. The impact of ISR on the cross-section is shown in
Fig.~\ref{linez}.

\bigskip
\noindent
\begin{minipage}[b]{.5\textwidth}
\paragraph{Backgrounds}: $\sigma^{\rm deconv}_f$ receives contributions from
several sources:
\begin{align}
\sigma^{\rm deconv}_f = \sigma^Z_f + \underbrace{\sigma^\gamma_f +
\sigma^{\gamma Z}_f + \sigma^{\rm box}_f}_{\sigma^{\rm bkgd}_f}
\end{align}
\end{minipage}
\hfill\raisebox{.5em}{%
\epsfig{figure=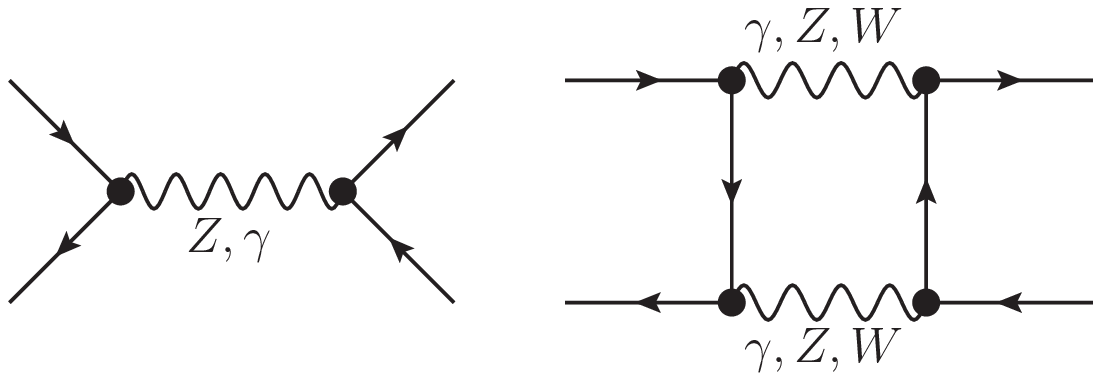, width=.4\textwidth, bb=140 600 460 710, clip=}}

\noindent
where $\sigma_Z$ stems from s-channel $Z$-boson exchange, $\sigma_\gamma$ from
s-channel photon exchange, $\sigma_{\gamma Z}$ from the inteference of these
two, and $\sigma_{\rm box}$ from box diagrams that involve the exchange of two
(or more) gauge bosons between the initial and final fermions.

Only $\sigma_Z$ has a Breit-Wigner resonance at $s\approx M_Z^2$, whereas the
remaining contributions in $\sigma^{\rm bkgd}_f$ are relatively suppressed.
For measurements near the $Z$ pole, the non-resonant terms in $\sigma^{\rm
bkgd}_f$ are typically subtracted, using their SM prediction
\cite{ALEPH:2005ab}.

\paragraph{Detector acceptance and cuts:} The measured cross-section is affected
by the capability of the detector to identify the final state $f\bar{f}$
particles, the presence of extra photon radiation in the detector, blind regions
of the detector, cuts to suppress backgrounds from fakes, etc. These effects are
typically evaluated using Monte-Carlo simulations.

\subsubsection{\boldmath $A_{\rm FB}$ at LHC}

In addition to $e^+e^-$ colliders, EWPOs can also be measured at hardon
colliders. However, a challenge arises when trying to determine the
forward-backward asymmetry at the LHC from the so-called \emph{Drell-Yan}
process $pp \to \ell^+\ell^-$ ($\ell=e,\mu$)\footnote{The achievable precision
for $\ell=\tau$ is strongly reduced since hadronic tau decay suffer from large
QCD backgrounds.}, since the initial state ($pp$) is symmetric and thus there
is no obvious distinction between the forward and backward directions.

%------------------------------------------------------------------------------
\begin{figure}
\centering
\begin{tabular}{c@{\hspace{3em}}c@{\hspace{3em}}c}
\epsfig{figure=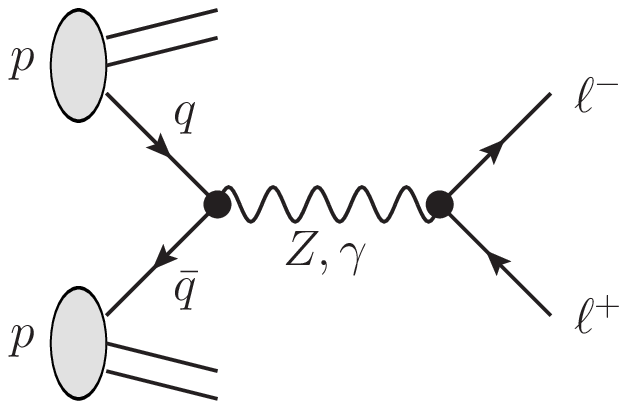, width=1.6in, bb=210 600 390 720} &
\raisebox{0.1in}{\epsfig{figure=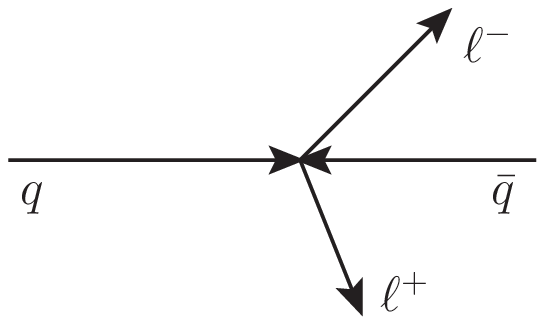, width=1.5in, bb=210 620 366 712}} &
\epsfig{figure=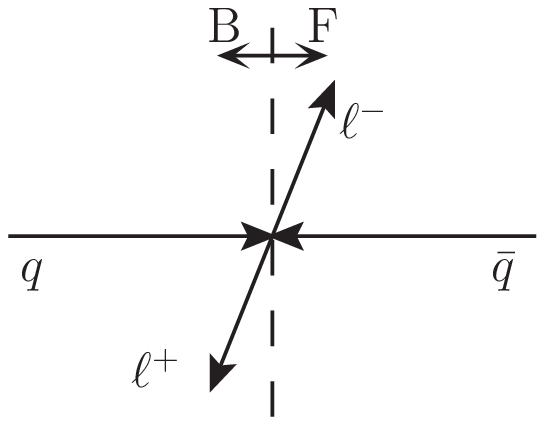, width=1.5in, bb=210 586 366 706} \\ 
(a) & (b) lab frame & (c) CoM frame
\end{tabular}
\caption{Drell-Yan process at LHC: (a) Leading Feynman diagram; (b) kinematics
in the lab frame, and (c) in the center-of-mass frame. The direction of the
boost from the center-of-mass to the lab frame is taken at the forward direction
to define $A_{\rm FB}$.}
\label{afbpp}
\end{figure}
%------------------------------------------------------------------------------
However, the leading partonic process consists of an asymmetric quark-antiquark
pair, see Fig.~\ref{afbpp}~(a). On average, the quark momentum is expected to be
larger than the antiquark momentum, since the quark may be a valence parton of the proton,
whereas the antiquark necessarily stems from the sea parton distribution.
Therefore, the final-state $\ell^+\ell^-$ will typically be boosted in the
direction of the incoming quark, Fig.~\ref{afbpp}~(b). To evaluate $A_{\rm FB}$,
the event must be tranformed to the center-of-mass frame, but one can use the
boost direction of the event in the lab frame to define the forward direction
for the asymmtry, Fig.~\ref{afbpp}~(c).

Given the large cross-section for $Z$-boson production at the LHC, there is the
potential to perfrom high-precision measurements of $A_{\rm FB}$ at the ATLAS
and CMS experiments. Nevertheless, the achievable precision is limited by systematic
effects:
\begin{itemize}
\item The overall boost direction of the event is not a perfect proxy for the
direction of the incident quark. To evaluate how often the forward and backward
hemispheres get incorrectly assigned, precise parton distribution functions
(PDFs) are needed. Thus the measurement precision for $A_{\rm FB}$ is limited by
the PDF errors.
\item Drell-Yan production receives large QCD corrections from gluon exchange
among the initial-state $q\bar{q}$ system. Recently, the NNNLO corrections have
been computed \cite{Duhr:2020seh}, but the error from unknown higher-order QCD
contribution is still not negligible.
\end{itemize}

\bigskip\noindent
Let us conclude this section by highlighting some examples of EWPO measurements.
The best measurement of the total $Z$ width has been obtained at LEP
\cite{ALEPH:2005ab}:
\begin{align}
\Gamma_Z &= 2495.5 \pm 2.3 \,\text{MeV}  && \text{(LEP)}
\end{align}
For the leptonic effective weak mixing angle, a number different
measurements of left-right and forward-backward asymmetries at lepton and hadron
colliders are similarly competitive
\cite{ALEPH:2005ab,Aaltonen:2018dxj,ATLAS-CONF-2018-037}:
\begin{align}
\seff{\ell} =\; &0.23098 \pm 0.00026 && (A_{\rm LR} \text{ @ SLD}) \notag \\
&0.23221 \pm 0.00029 && (A_{\rm FB}^b \text{ @ LEP}) \notag \\
&0.23148 \pm 0.00033 && (A_{\rm FB}^{e,\mu} \text{ @ TeVatron}) \notag \\
&0.23140 \pm 0.00036 && (A_{\rm FB}^{e,\mu} \text{ @ ATLAS})
\end{align}

%%%%%%%%%%%%%%%%%%%%%%%%%%%%%%%%%%%%%%%%%%%%%%%%%%%%%%%%%%%%%%%%%%%%%%%%%%%%%%%

\subsection{W-boson mass}

The $W$-boson mass is typically not considered an input parameters, since it can
be computed from the Fermi constant $G_\mu$, using eq.~\eqref{GmuSM}. Together
with $g=e/\sw = e/\sqrt{1-M_W^2/M_Z^2}$, this equation can be solved for
\begin{align}
M_W^2 = M_Z^2\biggl[\frac{1}{2} + \sqrt{\frac{1}{4} -
 \frac{\alpha\pi}{\sqrt{2}G_\mu M_Z^2}(1+\Delta r)}\biggr] \label{MWfromGmu}
\end{align}
$\Delta r$ in general depends on all parameters in the SM, including $M_W$, so
that \eqref{MWfromGmu} needs to be solved recursively.

\medskip\noindent
This prediction of $M_W$ can be compared to direct measurements. Currently, the
most precise determination of the $W$ mass is from hadron collider experiments,
using the process $pp \to \ell^\pm  \stackrel{\!\!\text{\tiny (}-\text{\tiny
)}\!\!}\nu_\ell$, which proceeds through an s-channel $W$-boson (at tree-level).
The $W$-boson mass thus corresponds to a peak in the invariant mass distribution
of the final state lepton-neutrino system:
\begin{align}
m_{\rm inv} = \sqrt{(p_\ell+p_\nu)^2} \approx \sqrt{2|\vec{p}_\ell||\vec{p}_\nu|
 - 2\vec{p}_\ell\cdot\vec{p}_\nu}
\end{align}
where in the last step we have neglected the masses of the lepton and neutrino.
The transverse component (perpendicular to the beam axis) of $\vec{p}_\nu$ can
be reconstructed by using momentum conservation, $\vec{p}_{\nu,\rm T} =
-\vec{p}_{\ell,\rm T}-\vec{p}_{X,\rm T}$, where $X$ are any jets or other
particles stemming from the proton remnants. However, since only a fraction of
the momenta of the incoming protons is transferred to the $W$ boson, the total
longitudinal momentum of the event is unknown, and thus one cannot reconstruct
$\vec{p}_{\nu,\rm L}$.

\begin{figure}
\centering
\epsfig{figure=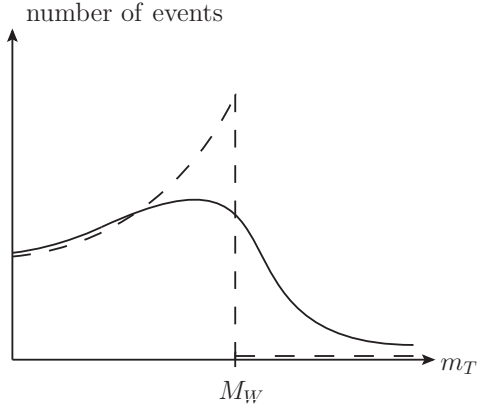, width=2.5in, bb=160 485 420 708}
\caption{(a) Sketch of the transverse mass distribution for $W$-boson production at
hadron colliders, for a perfect detectors and without width effects (dashed),
and including detector smearing and width effects (solid).}
\label{mtdist}
\end{figure}

Instead of the invariant mass distribution, one can utilize the \emph{transverse
mass}
\begin{align}
m_{\rm T} \equiv \sqrt{2|\vec{p}_{\ell,\rm T}||\vec{p}_{\nu,\rm T}|
 - 2\vec{p}_{\ell,\rm T}\cdot\vec{p}_{\nu,\rm T}}
\end{align}
One can straightforwardly show that $m_{\rm T} \leq m_{\rm inv}$. When
neglecting the $W$ width and assuming a perfect detector, $M_W$ thus corresponds
to the endpoint of the $m_T$ distribution. In reality, finite width effects and
detector smearing lead to a washed-out endpoint \cite{Smith:1983aa}, see
Fig.~\ref{mtdist}. Therefore, careful modeling of
detector effects and photon radiation is required for a precision measurment of
$M_W$ from the $m_T$ distribution.
%------------------------------------------------------------------------------
%------------------------------------------------------------------------------

\medskip\noindent
At lepton colliders, the $W$-mass can be measured from the invariant mass
distribution in the processes $e^+e^- \to W^+W^- \to qqqq$ and $e^+e^- \to
W^+W^- \to qq\ell\nu$. The reconstruction of both the transverse and
longitudinal components the neutrino momentum is possible here, since there is
no ambiguity due to the momentum carried away by the proton remnants.

Alternatively, one may measure $M_W$ from a threshold scan, by measuring the
cross-section for $e^+e^- \to W^+W^-$ at a few center-of-mass energies near
$2M_W$. Since the cross-section near threshold is small, this approach requires 
a large amount of luminosity. With the available statistics at LEP, the
achievable precision was rather low \cite{Schael:2013ita}. 

For the theoretical description of the cross-section as a function of $\sqrt{s}$
near threshold, one needs to compute the full process $e^+e^- \to
qqqq/qq\ell\nu$, since contributions where the $W$-bosons are off-shell are
important for $\sqrt{s} \lesssim 2M_W$. In fact, in this regime diagrams without
a $W^+W^-$ pair also contribute significantly. The currently most accurate
calculation includes full NLO corrections to the $e^+e^- \to 4f$ process
\cite{Denner:2005fg}.

The most precise available $M_W$ measurements are
\cite{Schael:2013ita,Aaltonen:2013iut,Aaboud:2017svj} 
\begin{align}
M_W =\; &80.376\pm0.033\;\text{GeV} && (m_{\rm inv} \text{ @ LEP}) \notag \\
&80.387\pm0.016\;\text{GeV} && (m_{\rm T} \text{ @ TeVatron}) \notag \\
&80.370\pm 0.019\;\text{GeV} && (m_{\rm T} \text{ @ ATLAS})
\end{align}

%%%%%%%%%%%%%%%%%%%%%%%%%%%%%%%%%%%%%%%%%%%%%%%%%%%%%%%%%%%%%%%%%%%%%%%%%%%%%%%

\subsection{Future \boldmath $e^+e^-$ colliders}

The experimental precision of EWPOs could be significantly improved at an
$e^+e^-$ collider with much larger luminosity than LEP or SLD. Such machines are
proposed primarily for the purpose of detailed measurements of Higgs boson
properties, but they could also perform electroweak measurements at $\sqrt{s}
\sim M_Z$ and $\sqrt{s} \sim 2M_W$. The FCC-ee \cite{Abada:2019zxq} and CEPC
\cite{CEPCStudyGroup:2018ghi} concepts are based on circular colliders, where
the ILC concept \cite{Baer:2013cma,Bambade:2019fyw} envisions a linear setup.
The baseline run scenario for ILC does not include any runs on the $Z$ pole and
near the $WW$ threshold, but it can study electroweak physics at a higher energy
$\sqrt{s} \sim 250$~GeV by using the \emph{radiative return} method, where
radiation of initial-state photons results in a lower effective center-of-mass
energy [see eq.~\eqref{isr}].

\medskip\noindent
The following table illustrates the expected improved precision for a few
selected EWPOs:

\begin{center}
\begin{tabular}{|l|cccc|}
\hline
 & today & FCC-ee & CEPC & ILC \\
\hline
$\Gamma_Z$ [MeV] & 2.3 & 0.1 & 0.5 & -- \\
$\seff{\ell}$ [$10^{-5}$] & 13 & 0.5 & $<1$ & $\sim 2$ \\
$M_W$ [MeV] & 12 & $\lesssim 1$ & $\sim 1$ & 2.4 \\
\hline
\end{tabular}
\end{center}

%%%%%%%%%%%%%%%%%%%%%%%%%%%%%%%%%%%%%%%%%%%%%%%%%%%%%%%%%%%%%%%%%%%%%%%%%%%%%%%

\subsection{Low-Energy EWPOs}
\label{lewpo}

Electroweak physics can also be studies with precision experiments performed at
lower energies, where the $W$ and $Z$ bosons appear only as virtual particles.
An overview of some variety of such experiments can be found in
Ref.~\cite{Erler:2013xha}. In the following, two types of such electroweak
precision tests will be briefly described.

\bigskip\noindent
\begin{minipage}{.72\textwidth}
\paragraph{Polarized electron scattering:} A beam of left- or right-handed $e^-$
is scattered off target particles $X$, where $X$ could be electrons, protons,
deuterons, or heavier nuclei. If $X$ is a hadronic or nuclear target, it is
advantageous to choose the kinematics such that the momentum transfer is small,
$q^2 \ll m_p^2$, so that the proton (or nucleus) can be regarded as
approximately pointlike.
\end{minipage}
\hfill\raisebox{-3em}{%
\epsfig{figure=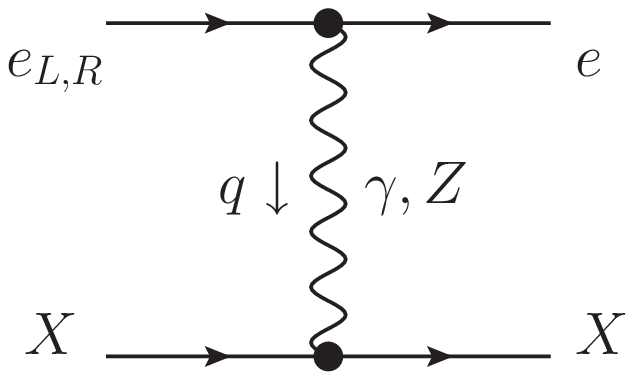, width=.23\textwidth, bb=200 610 384 718, clip=}}

\medskip
While the cross-section overall is strongly dominated by t-channel photon
exchange, one can probe electroweak physics through the left-right asymmetry
\begin{align}
A_{\rm LR} &= \frac{\sigma_{\rm L}-\sigma_{\rm R}}{\sigma_{\rm L}+\sigma_{\rm R}}
\intertext{For electron-proton scattering in the limit $q^2 \ll m_p^2$, this asymmetry is
given by, at tree-level,}
A_{\rm LR}^{ep} &\approx \frac{G_\mu(-q^2)}{4\sqrt{2}\pi \alpha}(1-4\sw^2)
\end{align}
Thus a measurement of $A_{\rm LR}^{ep}$ can be used to determine the weak mixing
angle.

Higher-order radiative corrections can be accounted for by replacing the
on-shell weak mixing angle $\sw$ with the effective weak mixing angle $\seff{}$, and by
including additional correction factors:
\begin{align}
1-4\sw^2 \quad\to\quad 1-4\kappa \seff{\ell} + \Delta Q
\end{align}
\begin{minipage}{.72\textwidth}
$\kappa$ includes large corrections from the $\gamma$--$Z$ mixing self-energy,
which are enhanced by large logarithms:
\begin{align}
\kappa \approx 1 - \frac{\cw}{12\pi^2\sw}\sum_f v_f (eQ_f) \ln
\frac{m_f^2}{M_Z^2}
\end{align}
Similar to the logarithms in the charge renormalization, eq.~\eqref{sigmap},
these are ill-defined for light quarks, $f=u,d,s$.
\end{minipage}
\hfill\raisebox{-3em}{%
\epsfig{figure=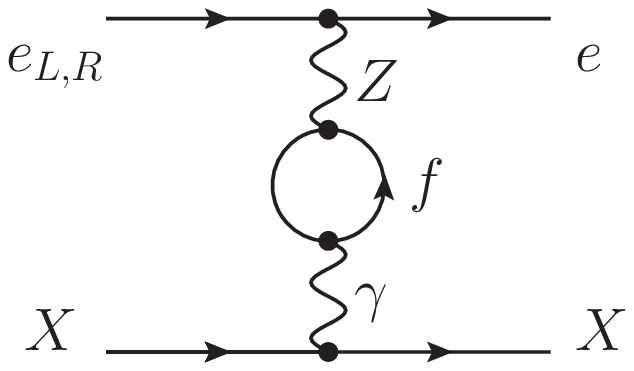, width=.23\textwidth, bb=200 610 384 718, clip=}}

Similar to what is done for $\Delta\alpha$, one may try to extract the hadronic
corrections to $\kappa$ from data for $R(s) = \frac{\sigma[e^+e^- \to
\text{hadrons}]}{\sigma[e^+e^- \to \mu^+\mu^-]}$ using a dispersion integral.
However, this requires additional assumptions in this case, such as SU(3)$_{\rm
u,d,s}$ flavor symmetry
\cite{Wetzel:1981vt,Marciano:1983wwa,Jegerlehner:1985gq,Jegerlehner:2017zsb},
because the $Z$ couplings have a different dependence on the fermion flavor that
$\gamma$ couplings.

Alternative, the leading hadronic effects can be absorbed into a running \msbar\
weak mixing angle \cite{Erler:2004in,Erler:2017knj},
\begin{align}
\kappa\seff{\ell} \approx \sin^2 \overline{\theta}(\mu^2=-q^2)
\equiv
\frac{\overline{g}'^2(\mu)}{\overline{g}^2(\mu)+\overline{g}'^2(\mu)}
\bigg|_{\mu^2=-q^2}
\end{align}
where the bar above an expression denotes that this quantity is defined in the
\msbar\ scheme.

The following table lists some of the current and near-future electron-electron
and electro-proton scattering experiments, together with their precision in
measuring the weak mixing angle
\cite{Anthony:2005pm,Androic:2018kni,Benesch:2014bas,Becker:2018ggl}: 
\begin{center}
\begin{tabular}{|l|cc|}
\hline
 & $ee$ & $ep$ \\
\hline
current & E158 (0.5\%) & Qweak (0.5\%) \\
future & MOLLER (0.1\%) & P2 (0.1\%) \\
\hline
\end{tabular}
\end{center}
The anticipated precision of the future MOLLER and P2 experiments will be
comparable to the combined $Z$-pole analysis from LEP/SLC, but in an entirely
different setup at low energies, with different sources of experimental and
theoretical systematic errors.
A more
comprehensive exposition of these types of experiments can be found $e.\,g.$ in
Ref.~\cite{Kumar:2013yoa}.

\paragraph{Muon anomalous magnetic moment:} Charged fermions have a magnetic
moment with the magnitude $\frac{eQ_f}{2m_f}g_f$, where $g_f$ is called the
Land\'e factor. At tree-level ($e.\,g.$ from the Dirac equation) $g_f=2$.
However, the value of $g_f$ gets modified through radiative corrections,
generating an anomalous magnetic moment $a_f = (g_f-2)/2 \neq 0$.

In the following we focus on the anomalous magnetic moment of leptons
\cite{Jegerlehner:2009ry,Jegerlehner:2018zrj}. The main
contribution to $a_\ell$ stems from QED, which has been computed up to ${\cal
O}(\alpha^5)$, see eq.~\eqref{aeth}.

Electroweak and hadronic corrections are suppressed by powers of
$m_\ell^2/M_W^2$ and $m_\ell^2/\Lambda_{\rm QCD}^2$, respectively. 
Thus they are negligble for the electron magnetic moment, but they become
imporant for $\ell = \mu$. The hadronic corrections are relatively large and are
typically extracted from data for $R(s) = \frac{\sigma[e^+e^- \to
\text{hadrons}]}{\sigma[e^+e^- \to \mu^+\mu^-]}$ \cite{Jegerlehner:2017lbd,Davier:2019can,Keshavarzi:2019abf}. The experimental error of this
data is the dominant uncertainty in the theoretical prediction of $a_\mu$.
Efforts to compute the hadronic corrections using lattice QCD have made a lot
of progress recently \cite{Meyer:2018til,Borsanyi:2020mff}.

The electroweak effects are rather small,
\begin{align}
a_\mu^{\rm EW} = \frac{g^2}{16\pi^2}\,\frac{m_\mu^2}{M_W^2} \times {\cal O}(1)
\sim 1.5\times 10^{-9} \label{amuEW}
\end{align}
but need to be taken into account given the experimental precision for the
measurement of $a_\mu$. The most precise experimental value is from the g--2
experiment at BNL \cite{Bennett:2004pv}, which yielded
\begin{align}
a_\mu^{\rm exp} = (11\,659\,208.0 \pm 6.3) \times 10^{-10}
\end{align}
which differs from the SM prediction \cite{Zyla:2020zbs}
\begin{align}
a_\mu^{\rm exp} = (11\,659\,184.6 \pm 4.7) \times 10^{-10}
\end{align}
by more than 3 standard deviations. An ongoing experiment at FNAL aims to
improve the precision of $a_\mu^{\rm exp}$ by a factor 4
\cite{Grange:2015fou}.

\paragraph{Exercise:} The electroweak corrections in \eqref{amuEW} are
proportional to $m_\mu^2$, and most corrections from BSM physics would have the
same proportionality. One power of $m_\mu$ stems from the fact that the magnetic
moment coupling ${\cal L} \supset \text{const.}\times \overline{\psi} \sigma_{\mu\nu} F^{\mu\nu}
\psi$ involves a derivative inside the field strength tensor and thus is
proportional to the overall energy scale of the process. Where does the other
power of $m_\mu$ comes from? Can it be replaced by something else in some new
physics model?
\label{alm}

%%%%%%%%%%%%%%%%%%%%%%%%%%%%%%%%%%%%%%%%%%%%%%%%%%%%%%%%%%%%%%%%%%%%%%%%%%%%%%%

\section{Tests of the Standard Model and Physics Beyond the Standard Model}
\label{bsm}

\subsection{Standard Model predictions}

The consistency and accuracy of the SM as a description of electroweak physics
can be tested by comparing experimental data for EWPOs with theoretical
predictions, where the latter care computed within the SM in as a function of a
set of input parameters. All the EWPOs discussed in the previous section can be
used for this purpose: $\Gamma_Z$, $\sigma_f^0$, $\seff{f}$, $M_W$ (predicted
from $G_\mu$), $a_\mu$, etc.

Owing to the precision of the available experimental data, higher-order
corrections need to be included in this comparison. For all EWPOs listed above,
complete two-loop corrections are known, as well as some partial higher-order
contributions, in particular from QED and QCD effects (see
Refs.~\cite{Freitas:2016sty,Dubovyk:2018rlg,Dubovyk:2019szj,Awramik:2003rn,Gnendiger:2013pva,Czarnecki:2002nt}
and references therein). While the one-loop corrections can be evaluated
analytically, with logarithms and dilogathrims appearing in the final result
\cite{Denner:1991kt}, the is in general not the case at the two-loop level and beyond.
Instead one needs to result to either approximations or numerical methods. The
numerical approaches can be divided into two groups:
\begin{itemize}
\item General techniques that can in principle be applied to problems with any
number of loops, external legs and types of particles. The best-known approach
in this category is sector decomposition \cite{Binoth:2000ps}, which allows one
to extract all UV and IR singularities with an algorithm that can be implemented
in computer programs and then integrate the coefficients of the singularities
and the finite remainder numerically
\cite{Bogner:2007cr,Borowka:2015mxa,Borowka:2017idc,Smirnov:2015mct}. Another
approach, which is not fully general but works for many two- and three-loop
applications, is based on Mellin-Barnes representations
\cite{Czakon:2005rk,Dubovyk:2016ocz}. The disadvantage of these techniques is
their relatively large need of computing resources for the evaluation of
multi-dimensional numerical integrals that are slowly converging.
\item A range of numerical methods have been tailored for a particular type of
problem, $i.\,e.$ self-energy or vertex integrals of a certain loop order. While
limited in scope, these approaches tend to produce numerical integrals of lower
dimensionality and more favorable convergence behavior than the general
techniques. A review of can be found in Ref.~\cite{Freitas:2016sty}.
\end{itemize}
It is instructive to look at some of the leading effects of the radiative
corrections. For this purpose, let us consider the corrections to
the Fermi constant, see eq.~\eqref{GmuSM}, and to the effective weak mixing
angle, see eq.~\eqref{sweff}. They may be written as
\begin{align}
&\frac{G_\mu}{\sqrt{2}} = \frac{g^2}{8M_W^2}(1+\Delta r),
& \Delta r &= \Delta\alpha - \frac{\cw^2}{\sw^2}\Delta\rho + \Delta r_{\rm rem},
\\
&\seff{f} = \sw^2(1+\Delta\kappa),
& \Delta\kappa &= \frac{\cw^2}{\sw^2}\Delta\rho + \Delta \kappa_{\rm rem}
\end{align}
Here $\Delta r$ and $\Delta\kappa$ include all higher-order corrections. Two
leading contributions can be identified:

The shift in the fine structure constant, $\Delta\alpha$, has already been
discussed on page~\pageref{deltaalpha}. It receives numerically
comparable contributions from both leptonic and hadronic loops, which add up to
\begin{align}
\Delta\alpha = \Delta\alpha_{\rm lept} + \Delta\alpha_{\rm had} \approx 6\%
\end{align}
The numerical enhancement stems from the logarithmic dependence on light fermion
masses, see eq.~\eqref{sigmap}.

On the other hand, $\Delta\rho$ contains contributions that are proportional to
the Yukawa couplings of fermions inside the loop, where the top Yukawa $y_t
\approx 1$ dominates, whereas all other fermions are negligible:
\begin{align}
\Delta\rho = \frac{3y_t^2}{32\pi^2} + 
\underbrace{...}_{\text{fermions other than the top}\hspace{-8em}}\hspace{4em}
\end{align}
It appears in $\Delta r$ and $\Delta\kappa$ in the combination
$\frac{\cw^2}{\sw^2}\Delta\rho \approx 3\%$. The remaining corrections are
numerically smaller: $\Delta r_{\rm rem}$, $\Delta\kappa_{\rm rem} \lesssim
1\%$.

When comparing $G_\mu$, $M_W$ and $\seff{f}$ to data, the dominant effect of
$\Delta\rho$ leads to a relatively precise indirect determination of the top
mass, $m_t = y_t v/\sqrt{2} = 176.3 \pm 1.9\,$GeV, which agrees reasonably well
with the direct measurement from LHC and Tevatron, $m_t^{\rm exp} = 172.9 \pm
0.3\,$GeV (see section 10 of Ref.~\cite{Zyla:2020zbs}).

On the other hand, the indirect determination of $M_H$ from electroweak
precision data is much less accurate \cite{Zyla:2020zbs}, since the $M_H$ only
appears in the small terms $\Delta r_{\rm rem}$, $\Delta\kappa_{\rm rem}$, and
the functional dependence on $M_H$ is only logarithmic.

The numerically large quadratic dependence on $y_t$ in $\Delta\rho$ can be
explained through the breaking of \emph{custodial symmetry}. This is a symmetry
of the Higgs potential, which can be most easily seen by re-writing the Higgs
field as a matrix. Since the complex Higgs field $\phi = \begin{pmatrix} \phi^+
\\ \phi^0 \end{pmatrix}$ has four physical degrees of freedom, one can arrange
these four components into a matrix,
\begin{align}
\Omega = \begin{pmatrix} \phi^{0*} & \phi^+ \\ \phi^- & \phi^0 \end{pmatrix}
\end{align}
where $\phi^{0*}$ and $\phi^-$ are the conjugate fields of $\phi^0$ and
$\phi^+$, respectively. Then the scalar potential becomes
\begin{align}
V = -\mu^2|\phi|^2 + \lambda|\phi|^4 = 
 - \frac{\mu^2}{2}\,\text{Tr}\{\Omega^\dagger\Omega\}
 + \frac{\lambda}{4}\bigl(\text{Tr}\{\Omega^\dagger\Omega\}\bigr)^2
\end{align}
In this form, one can see that $V$ is manifestly invariant under transformations
\begin{align}
\Omega \to L\Omega R^\dagger, 
\qquad L \in \text{SU(2)}_L, \quad R \in \text{SU(2)}_R \label{lrsymm}
\end{align}
where $L,R$ are unitary SU(2) matrices. Since $L$ and $R$ can be independent of
each other, they are part of two separate symmetry groups, labeled
SU(2)$_{L,R}$. SU(2)$_L$ is the usual weak symmetry group.

When $\phi$ obtains a vev, $\langle \Omega \rangle = \begin{pmatrix} v & 0 \\ 0 &
v \end{pmatrix}$, the symmetry \eqref{lrsymm} will be broken, but $\langle
\Omega \rangle$ is still invariant under a symmetry sub-group where $L=R\equiv V$:
\begin{align}
\langle\Omega\rangle \to V\langle\Omega\rangle V^\dagger, 
\qquad V \in \text{SU(2)}_{\rm diag}
\end{align}
$\text{SU(2)}_{\rm diag}$ is called the ``custodial symmetry'' group. The SM
Higgs potential, Higgs vev, and weak and QCD gauge interactions are invariant
under it, but not the Yukawa couplings. As an example, let us consider the
Yukawa couplings of the top and bottom quarks,
\begin{align}
{\cal L}_{\rm Yuk,tb} = -y_t \overline{Q}_{3L} \tilde{\phi}\, t_R - y_b
 \overline{Q}_{3L} \phi\, b_R + \text{h.c.}, \qquad \overline{Q}_{3L} = \begin{pmatrix} t_L \\
 b_L \end{pmatrix}
\end{align}
Here $\tilde{\phi} = C\phi^*$, and $C = i\sigma^2$ is the charge conjugation matrix.
If $y_t$ and $y_b$ were equal, $y_t=y_b \equiv y$, this could be re-written as
\begin{align}
{\cal L}_{\rm Yuk,tb} = -y\, \overline{Q}_{3L} \Omega\, Q_{3R} + \text{h.c.}, 
\qquad \overline{Q}_{3R} = \begin{pmatrix} t_R \\ b_R \end{pmatrix}
\end{align}
which would be invariant under $\text{SU(2)}_{\rm diag}$ if the quarks doublets
transform as $Q_{L,R} \to VQ_{L,R}$. 

However, the fact that $y_t \neq y_b$ leads to breaking of $\text{SU(2)}_{\rm
diag}$. Any breaking effect must be proportional to some power of $(y_t - y_b)^2
\approx y_t^2$,
so that is vanishes in the limit where $\text{SU(2)}_{\rm diag}$ is restored.
This is the origin of the effect in $\Delta\rho$ proportional to $y_t^2$.

Note that $\text{SU(2)}_{\rm diag}$ is also broken by the hypercharge gauge
coupling, but the numerical impact of that in EWPOs is smaller.

%%%%%%%%%%%%%%%%%%%%%%%%%%%%%%%%%%%%%%%%%%%%%%%%%%%%%%%%%%%%%%%%%%%%%%%%%%%%%%%

\subsection{Constraints on Physics Beyond the Standard Model}

A global fit to all relevant EWPOs yields good agreement within the SM
\cite{Zyla:2020zbs}, and there is no obvious hint for BSM physics, except for
the discrepancy in the $a_\mu$ (see section~\ref{lewpo}). Thus the data can be
used to set constraints on new physics models. Based on the discussion from the
previous subsection, one can already conclude the models with new sources of
custodial symmetry breaking will be severely bounded by electroweak precision
data.

If one assumes that the new degree of freedom beyond the SM are heavy compared
to the electroweak scale, one BSM effects in EWPOs can be parametrized in a
model-independent way by adding higher-dimensional operators to the theory. This
framework is often referred to as SMEFT (SM Effective Field Theory).
The leading contribution for EWPOs stems from dimension-6 operators,
\begin{align}
{\cal L} = {\cal L}_{\rm SM} + \sum_i \frac{C_i}{\Lambda^2}{\cal O}_i,
\end{align}
where $\Lambda \gg v$ is the mass scale of the BSM physics (typically the
smallest BSM mass if there is a more complex particle spectrum). The complete
list of dimension-6 operators ${\cal O}_i$ can be found $e.\,g.$ in
Ref.~\cite{Grzadkowski:2010es}. The values of the Wilson coefficients depend on
the underlying BSM physics, and they can be computed in terms of the parameters
of a specific hypothetical model (a procedure called ``matching'').

By comparing data to predictions for EWPOs within SMEFT, constraints on the
Wilson coefficients of a subset of operators can be derived. The subset that
EWPOs are sensitive to includes operators that modify gauge-boson--fermion
couplings, Higgs-boson--gauge-boson interactions, and certain four-fermion
interactions. In principle, these constraints can be derived in a
model-independent fashion, but since there are more operators than independent
observables, certain assumptions are typically imposed. For example, one may
assume \emph{flavor universality}, which means that the Wilson coefficients for
operators involving fermions are the same for all three fermion generations.

A more detailed description of SMEFT and its applications can be found in the
lectures on ``Standard Model Effective Field Theories'' in this school
\cite{martinlect}.

\bigskip\noindent
In the following, we will instead focus on BSM models where the scale of new
physics is lower, $\Lambda \lesssim v$, and thus the SMEFT is not applicable. In
the spirit of the school's theme, ``The Obscure Universe: Neutrinos and Other
Dark Matters,'' the focus is on examples that relate to neutrino and dark
matter physics.

%%%%%%%%%%%%%%%%%%%%%%%%%%%%%%%%%%%%%%%%%%%%%%%%%%%%%%%%%%%%%%%%%%%%%%%%%%%%%%%

\subsection{Neutrino Counting}

Decays of the $Z$ boson to neutrinos are invisible to collider detectors.
However, the existence of this decay channel can be probed by determining the
total width $\Gamma_Z$ from a fit to the Breit-Wigner lineshape and subtracting
the rates for all visible decay channels from it,
\begin{align}
\Gamma_Z = 3\,\Gamma_\ell + N_\nu \Gamma_\nu + \Gamma_{\rm had}
\end{align}
Here the masses of charged leptons and neutrinos have been neglected, so that
$\Gamma_e = \Gamma_\mu = \Gamma_\tau \equiv \Gamma_\ell$ and $\Gamma_{\nu_e} =
\Gamma_{\nu_\mu} = \Gamma_{\nu_\tau} \equiv \Gamma_\nu$. $N_\nu$ is the number
of neutrino species ($N_\nu=3$ in the SM).

$\Gamma_\ell$ and $\Gamma_\nu$ are not observables by themselves. However, they
can be related to observables as follows \cite{ALEPH:2005ab,Janot:2019oyi}:
\begin{align}
N_\nu &= \biggl[\biggl(\frac{12\pi}{M_Z^2}\,\frac{R_\ell}{\sigma^0_{\rm had}}
	\biggr)^2 - R_\ell - 3\biggr] \, \frac{\Gamma_\ell}{\Gamma_\nu}
\intertext{Here}
\sigma^0_{\rm had} &= \sigma_{e^+e^- \to \rm had}(s{=}M_Z^2)
 = \frac{12\pi}{M_Z^2}\,\frac{\Gamma_\ell\Gamma_{\rm had}}{\Gamma_Z^2}, \\
R_\ell &= \frac{\Gamma_{\rm had}}{\Gamma_\ell}
 = \frac{\sigma^0_{\rm had}}{\sigma^0_\ell}
\end{align}
can be determined from data, and ``had'' refers to all hadronic final state
($i.\,e.$ summing over all quarks $q \neq t$ in the partonic picture). On the
other hand, $\Gamma_\ell/\Gamma_\nu$ is computed in the SM, but the result is
correct also in a variety of models with extended neutrino sectors. Using
measurements from LEP, one finds \cite{Janot:2019oyi}
\begin{align}
N_\nu = 2.996 \pm 0.007
\end{align}
in agreement with SM expectations.

If one assumes that any BSM neutrino is part of an SU(2)$_L$ doublet together
with a new charged lepton ($i.\,e.$ a fourth lepton family), 
an additional constraint follows from the
contribution of this lepton doublet to $\Delta\rho$:
\begin{align}
\Delta\rho_{\nu\ell4} = \frac{1}{32\pi^2} \biggl[\underbrace{y^2_{\ell 4} + y^2_{\nu 4}
 - \frac{4y^2_{\ell 4} y^2_{\nu 4}}{y^2_{\ell 4} - y^2_{\nu 4}}
 \ln \frac{y_{\ell 4}}{y_{\nu 4}}}_{\geq (y_{\ell 4}-y_{\nu 4})^2} \biggr]
\end{align}
Here $y_{\ell 4}$ and $y_{\nu 4}$ are the Yukawa couplings of the extra
charged lepton and extra neutrino, respectively.
The expression in [ ] can be shown to be bounded from below by $(y_{\ell
4}-y_{\nu 4})^2$. 

EWPO data puts a constraint on any new physics contributions to $\Delta\rho$,
leading to the bound (at 90\% confidence level) \cite{Zyla:2020zbs}
\begin{align}
|y_{\ell 4}-y_{\nu 4}| < 48\,\text{GeV} \label{lnudiff}
\end{align}
Extra charged leptons would be visible in particles detectors, of course.
Searches at LEP2 exclude the existence of any such particle with mass below 101
GeV \cite{Zyla:2020zbs}. Together with \eqref{lnudiff} this implies that a 4th
generation neutrino with $m_{\nu 4} < 50$~GeV is excluded.

At the same time, studies of the $H \to \gamma\gamma$ rate forbid the existence
of a 4th lepton family where both the $\ell_4$ and $\nu_4$ are heavy
\cite{Lenz:2013iha}, so that the combination of electroweak precision and Higgs data
fully rules out the existence of a sequential 4th fermion generation.

%%%%%%%%%%%%%%%%%%%%%%%%%%%%%%%%%%%%%%%%%%%%%%%%%%%%%%%%%%%%%%%%%%%%%%%%%%%%%%%

\subsection{Sterile Neutrinos}

The bounds in the previous subsection do not apply to new neutral fermions that
are singlets under SU(2)$_L$, $i.\,e.$ that do not (electro)weak interactions.
Such particles are called \emph{sterile neutrinos} or \emph{right-handed
neutrinos}, since they can form Yukawa couplings with the SM neutrinos.

Let us consider a model where two such sterile neutrinos are added, denoted
$N_R^1$ and $N_R^2$, with the interaction Lagrangian \cite{Antusch:2015mia}
\begin{align}
{\cal L} = {\cal L}_{\rm SM} + \sum_k i\overline{N}_R^k \slashed{\partial}
N_R^k
 - \Bigl[ \sum_\alpha Y_{\nu\alpha} \overline{L}_{\alpha L}^{}
 \tilde{\phi} N_R^1 - M\,\overline{N}_R^1 C N_R^2 + \text{h.c.}\Bigr ]
\end{align}
Here $L_{1L} = \begin{pmatrix} \nu_{eL} \\ e_L \end{pmatrix}$ etc.\ are the SM
lepton doublets and $C$ is again the charge conjugation matrix.

In the limit that $M$ is much larger than the observed light neutrino masses,
the mass eigenstates of the model are:
\begin{itemize}
\item A pseudo-Dirac sterile neutrino $N$ with mass $\approx M$. Here the term
``pseudo-Dirac'' is used for a pair of Majorana fields with nearly degenerate masses,
which behave like a single Dirac particle in some phenomenological contexts.
$N$ is mostly composed of $N_R^{1,2}$, with a small admixture of left-handed SM
neutrinos $\nu_{\alpha L}$, so that is has strongly suppressed couplings to other SM particles
and could have escaped detection until now.
\item Active Majorana neutrinos $\nu'_{e,\mu,\tau}$, which are mostly SM-like,
with a small admixture of $N_R^{1,2}$, where the mixing angle is approximately given
by $\theta_\alpha \approx \frac{Y_{\nu\alpha}v}{\sqrt{2}M}$.
\end{itemize}
Assuming that $M > v$, the main phenomenological effect of this model, compared
to the SM, are reduced couplings of the active neutrinos to gauge bosons.
\begin{itemize}
\item In muon decay, $\mu \to e \nu'_\mu \bar{\nu}'_e$, the relationship between
Fermi constant and SM parameters is modifieid according to
\begin{align}
\frac{G_\mu}{\sqrt{2}} = \frac{g^2}{8M_W^2}(1+\Delta r)(1-\theta_e^2)(1-\theta_\mu^2)
\end{align}
\item The invisible $Z$ decay rate is reduced,
\begin{align}
\Gamma_{Z \to \rm inv} = \Gamma_\nu^{\rm SM}\Bigl(N_\nu -
\sum_{\alpha,\beta}\theta_\alpha\theta_\beta\Bigr) \label{gaminvnu}
\end{align}
\end{itemize}
In the above formulae, $\sin\theta_\alpha$ and $\cos\theta_\alpha$ have been
expanded for $\theta_\alpha \ll 1$.
Comparing these expressions to electroweak precison data, one obtains the bounds
\cite{Antusch:2015mia}
\begin{align}
&& \theta^2_e,\theta^2_\mu &\lesssim 2\times 10^{-3}, &
 \theta^2_\tau &\lesssim 7\times 10^{-3} && \text{(today)} && \notag \\
&& &\lesssim 2\times 10^{-5}, &
 &\lesssim 10^{-3} && \text{(FCC-ee)} && \notag \\
&& &\lesssim 2\times 10^{-5}, &
 &\lesssim 3\times 10^{-3} && \text{(CEPC)} && 
\end{align}

\paragraph{Exercise:} Assuming a special scenario where 
$\theta_e=\theta_\mu=\theta_\tau\equiv \theta$, what bound on $\theta$ (at 95\%
C.L.) to you obtain from \eqref{gaminvnu}. Use numbers from section 10 in
Ref.~\cite{Zyla:2020zbs} for $\Gamma_{Z \to \rm inv}^{\rm exp}$ and
$\Gamma_\nu^{\rm SM}$.
\label{thlim}

%%%%%%%%%%%%%%%%%%%%%%%%%%%%%%%%%%%%%%%%%%%%%%%%%%%%%%%%%%%%%%%%%%%%%%%%%%%%%%%

\subsection{Dark Photon}

Dark photon models are extensions of the SM with an additional U(1) gauge boson,
$Z'$ that can kinetically mix with the hypercharge gauge boson (see
Ref.~\cite{Fabbrichesi:2020wbt} for a recent review). Let us furthermore
introduce a fermion $\chi$ as a dark matter (DM) candidate that couples to $Z'$
with coupling strength $g_D$. The Lagrangian is given by
\begin{align}
{\cal L} = {\cal L}_{\rm SM} - \frac{1}{4}Z'_{\mu\nu} Z'^{\mu\nu} +
 \frac{M_{Z'}^2}{2}Z'_\mu Z'^\mu + \overline{\chi} (i\slashed{\partial}
  +g_D \slashed{Z}'-m_\chi)\chi + \frac{\epsilon}{2\cw}Z'_{\mu\nu} B^{\mu\nu}
\end{align}
Here we have written an explicit mass term for $Z'$ for simplicity. In
a realistic model this mass would need to be generated through the Higgs or
St\"uckelberg mechanism, but the details are unimportant for the following
discussion.

The kinetic terms can be diagonalized can canonically normalized by transforming
$Z'$ and $B$ to the new fields $Z^\mu_{D,0}$ and $B^\mu_0$ according to
\begin{align}
\begin{pmatrix} Z^\mu_{D,0} \\ B^\mu_0 \end{pmatrix}
 \approx \begin{pmatrix} 1-2\epsilon^2/\cw^2 & 0 \\ -\epsilon/\cw & 1  \end{pmatrix}
 \begin{pmatrix} Z'^\mu \\ B^\mu \end{pmatrix} + {\cal O}(\epsilon^3) 
 \label{zdmix}
\end{align}
When expressing ${\cal L}$ in terms of the $Z^\mu_{D,0}$ and $B^\mu_0$, one can
see that the dark photon field $Z^\mu_{D,0}$ has ${\cal O}(\epsilon)$ 
couplings to the SM fermions.

After electroweak symmetry breaking, mass mixing between
$B_0^\mu$, $W_0^\mu$ and $Z^\mu_{D,0}$ produces the
observable photon and $Z$-boson, as well as the ``dark photon'' mass eigenstate
$Z_D$ with mass $M_{Z_D}$. Note that the mass mixing between $Z^\mu_{D,0}$ and the other fields is
also suppressed by $\epsilon$. As a result, the $Z$-boson mass is shifted by an
${\cal O}(\epsilon^2)$ contribution relative to the SM \cite{Curtin:2014cca},
\begin{align}
M_Z^2 \approx \frac{M_W^2}{\cw^2}\Bigl(1+\epsilon^2\frac{\sw^2}{\cw^2}\Bigr)
\end{align}
where $\sw$ and $\cw$ are the sine and cosine of the weak mixing angle defined
through the (tree-level) gauge-couplings, $\sw = g'/\sqrt{g^2+g'^2}$, $\cw =
g/\sqrt{g^2+g'^2}$.

The $Zff$ vector and axial-vector couplings, see eq.~\eqref{zcpl}, are
additionally modified through $Z$--$Z_D$ mass mixing, leading to \cite{Curtin:2014cca}
\begin{align}
v_f &\approx \frac{e}{2\sw\cw}\biggl[\Bigl(1-\frac{\alpha^2}{2}\Bigr)
\bigl(I_f^3 - 2\sw^2 Q_f\bigr) + \alpha\epsilon\frac{\sw^2}{\cw^2} \bigl(Q_f -
I^3_f\bigr)\biggr], \\
a_f &= v_f|_{Q_f \to 0}^{}
\intertext{where}
\alpha &= \frac{\epsilon\sw}{\cw(M^2_{Z'}/M^2_{Z,0}-1)}\,, \qquad M_{Z,0} = 
\frac{M_W}{\cw}
\end{align}
As a result, the predictions for all $Z$-pole EWPOs are modified, such as
$\Gamma_Z$, the $Z$ branching ratios, and $\seff{f}$.

Additionally, the dark photon also leads to a correction of the electron and
muon magnetic moments \cite{Pospelov:2008zw},
\begin{align}
\delta a_\ell = \frac{\alpha\epsilon^2}{8\pi}\,
F\Bigl(\frac{m^2_\ell}{M^2_{Z_D}}\Bigr) \label{alZd}
\end{align}
where $F(x)$ is a function which is $F(x) \approx 1$ for $x \gg 1$ and $F(x)
\approx \frac{2}{3}x$ for $x \ll 1$. For some range of $\epsilon$ and $M_{Z_D}$,
\eqref{alZd} can explain the $>3\sigma$ discrepancy of the muon magnetic moment,
see section~\ref{lewpo}. However, for very small values of $M_{Z_D}$ the
correction to $a_e$ also can become sizeable and this region of parameter space
is ruled out. 

%------------------------------------------------------------------------------
\begin{figure}[tb]
\centering
\epsfig{figure=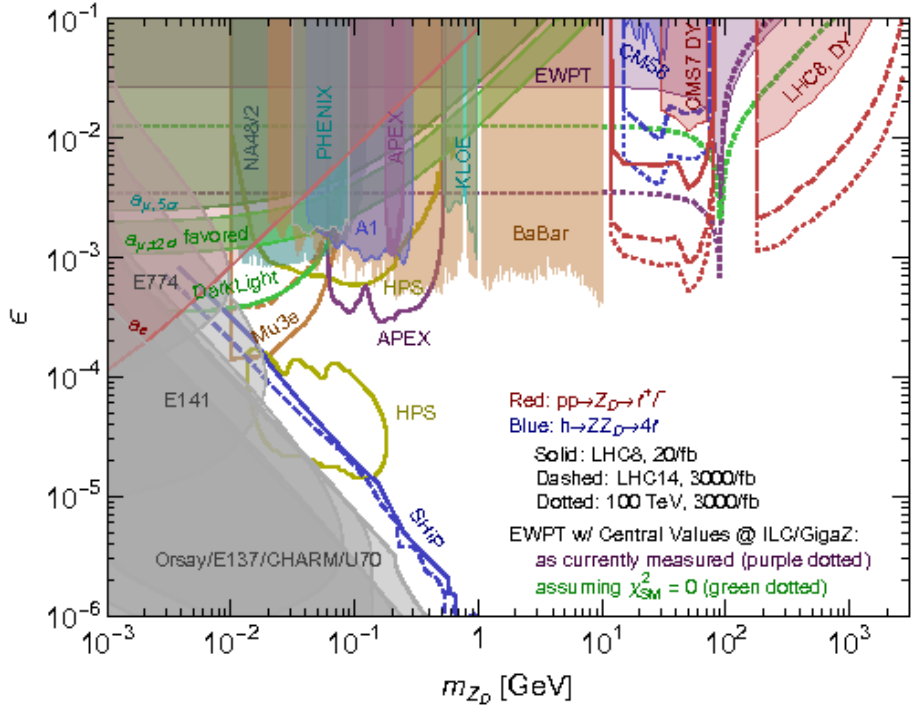, width=12cm}
\caption{Constraints on the parameter space of the dark photon model (figure
taken from Ref.~\cite{Curtin:2014cca}).
Electroweak precision constraints from $Z$-pole data are labeled ``EWPT'' (shaded
region for existing constraints and short-dashed lines for future constraints
obtainable at ILC). ``$a_{\mu,\pm 2\sigma}$ favored'' indicates the region that
would alleviate the muon magnetic moment discrepancy within 95\% confidence
level, while the region ``$a_{\mu,5\sigma}$'' is excluded because it would
worsen the discrepancy to the level of 5 standard deviations. The shaded region
labeled ``$a_e$'' is excluded from electron magnetic moment constraints. The
other shaded regions are excluded by direct searches for $Z_D$.}
\label{Zd_constr}
\end{figure}
%------------------------------------------------------------------------------
The constraints from $Z$-pole EWPOs and magnetic moments are
depicted in Fig.~\ref{Zd_constr}, together with bounds from direct searches for
$Z_D$ at various experiments. Note, however, that the direct search limits
assume that $Z_D$ only decays into SM particles. These bounds can be relaxed
when the invisible decay channel into DM particles, $Z_D \to
\chi\bar{\chi}$ is kinematically open ($m_\chi < M_{Z_D}/2$), whereas the bounds
from electroweak precision data are independent from such assumptions.

%%%%%%%%%%%%%%%%%%%%%%%%%%%%%%%%%%%%%%%%%%%%%%%%%%%%%%%%%%%%%%%%%%%%%%%%%%%%%%%

\bibliographystyle{JHEP}
\bibliography{stanmodel}

%%%%%%%%%%%%%%%%%%%%%%%%%%%%%%%%%%%%%%%%%%%%%%%%%%%%%%%%%%%%%%%%%%%%%%%%%%%%%%%

%\vspace{5em}

\appendix
\section{Answers to Exercise Questions}

Page~\pageref{se1a}:
\begin{align}
&\Sigma_T(k^2) = \frac{\alpha}{3\pi}\biggl[ 
 \frac{3(d/2-1)k^2+6m^2}{d-1}B_0(k^2,m^2,m^2) - \frac{4(d-2)}{d-1} A_0(m^2)
 \biggr] \notag \\
&\frac{\partial^2}{\partial k_\mu \partial k^\mu}f(k^2)
 = \frac{\partial}{\partial k_\mu}\biggl(\frac{\partial f}{\partial(k^2)}
 \underbrace{\frac{\partial(k^2)}{\partial k^\mu}}_{2k_\mu} \biggr) 
 = \frac{\partial f}{\partial(k^2)} 2d + 
 \frac{\partial^2 f}{\partial(k^2)^2} \underbrace{(2k_\mu)(2k^\mu)}_{0 \text{
 for } k^2=0} \notag \\
&\begin{aligned}
\frac{\partial}{\partial(k^2)}&B_0(k^2,m^2,m^2)\Big|_{k^2=0} \\
&=  \frac{1}{2d}\,\frac{\partial^2}{\partial k_\mu \partial k^\mu} \biggl[
 \int \frac{d^d q \; \omega_d}{[q^2-m^2][(q+k)^2-m^2]} \biggr]_{k^2=0}
 \hspace{7em} \biggl(\omega_d = \frac{(2\pi\mu)^{4-d}}{i\pi^2}\biggr) \\
&= \frac{1}{2d}\,\frac{\partial}{\partial k_\mu} \biggl[ \int
 \frac{d^d q \; \omega_d\;(-2)(q_\mu+k_\mu)}{[q^2-m^2][(q+k)^2-m^2]^2} 
 \biggr]_{k^2=0}\\
&= \frac{1}{2d}\biggl[ -
 \int \frac{d^d q \; \omega_d\; 2d}{[q^2-m^2][(q+k)^2-m^2]^2} +
 \int \frac{d^d q \; \omega_d\; 8(q+k)^2}{[q^2-m^2][(q+k)^2-m^2]^3}\biggr]_{k^2=0} \\
&=
\frac{1}{2d}\biggl[ (8-2d) \int \frac{d^d q \; \omega_d}{[q^2-m^2]^3}
 + 8m^2 \int \frac{d^d q \; \omega_d}{[q^2-m^2]^4}\biggr] \\
&= \frac{1}{2d}\bigl[ (8-2d) \tfrac{1}{2} A''_0(m^2) + 8m^2 \tfrac{1}{6} A'''_0(m^2) \bigr]
\end{aligned} \notag \displaybreak[0] \\[1ex]
\Rightarrow\quad &\Sigma'_T(0) = \frac{\alpha}{3\pi}\biggl[ 
 \frac{3(d/2-1)}{d-1}\underbrace{B_0(0,m^2,m^2)}_{A'_0(m^2)}
 + \frac{6m^2}{(d-1)2d}\bigl[ (4-d) A''_0(m^2) + \tfrac{4}{3}m^2 A'''_0(m^2) \bigr]
 \biggr] \notag \\
&A'_0(m^2) = \frac{2}{4-d} - \gamma_E - \ln\frac{m^2}{4\pi\mu^2},
\qquad A''_0(m^2) = -\frac{1}{m^2}, \qquad A'''_0(m^2) = \frac{1}{m^4} \notag \\
\Rightarrow\quad &\Sigma'_T(0) = \frac{\alpha}{3\pi}\biggl[ \frac{2}{4-d} -
\gamma_E - \ln\frac{m^2}{4\pi\mu^2} + {\cal O}(d-4) \biggr] \notag
\end{align}

\bigskip\noindent
Page~\pageref{dV}: 
\begin{align}
\delta V = [\text{terms in }\eqref{delV}] &+
 \frac{M_H^2}{2v}\Bigl(\frac{\delta v}{v} + \frac{\delta M_H^2}{M_H^2}\Bigr)
 \,h(h^2 + G_0^2 + 2G^+G^-) \notag \\
 &+ \frac{M_H^2}{8v^2}\Bigl(\frac{\delta v}{v} + \frac{\delta M_H^2}{M_H^2}\Bigr)
 \, (h^2 + G_0^2 + 2G^+G^-)^2 \notag
\end{align}

\bigskip\noindent
Page~\pageref{input}: 
\begin{itemize}
\item $\alpha$: protected by electromagnetic gauge symmetry;
\item $G_\mu$: new physics decouples in effective Fermi model (as long as
$m_{BSM} \gg m_\mu$);
\item $\alpha_{\rm s}$: protected by QCD gauge symmetry (lattice),
soft-collinear effective theory (event shapes, $\tau$ decays), or not at all
(EWPOs);
\item $m_t$: measurement based on kinematical feature (threshold);
\item $M_Z$: measurement based on kinematical feature (resonance).
\end{itemize}

\bigskip\noindent 
Page~\pageref{alm}: The magnetic moment interaction flips the
chirality of the fermion ($e.\,g.$ left- to right-handed). The only parameters
in the SM that break conservation of chirality are the Yukawa couplings (or,
equivalently, the fermion masses), and thus any chirality flip must be
proportional to $m_f$. A new physics model that has an additional source of
chirality breaking could change this behavior. For example, if one introduces a
second Higgs doublet that does not obtain any vev  (and thus does not contribute
to the muon mass) but has a Yukawa couplings $Y'_\mu$ with the muon, it could
generate a correction to $a_\mu$ proportional to $m_\mu Y'_\mu$ instead of
$m_\mu^2$. Since there are few other bounds on the value of $Y'_\mu$, this
corrections could be relatively large (even though a two-loop diagram would be
required).

\bigskip\noindent 
Page~\pageref{thlim}: According to Tab.~10.6 in \cite{Zyla:2020zbs}, 
$\Gamma_{Z \to \rm inv}^{\rm exp} = 498.9 \pm 2.5$~MeV and
$3\Gamma_\nu^{\rm SM} = 501.464$~MeV. The uncertainty in the latter is
negligible. Using 
$\Gamma_{Z \to \rm inv}^{\rm exp} = 3\Gamma_\nu^{\rm SM}(1-3\theta^2)$ and
assuming Gaussian error distribution, we then obtain $\theta^2 < 5\times
10^{-3}$ at 95\% C.L.

\end{document}